\documentclass[longbibliography,aps,prl,twocolumn,superscriptaddress,showkeys,amsfonts,amssymb,amsmath,floatfix]{revtex4-2}
\usepackage{inputenc}
\usepackage{amsmath,amsbsy,amsfonts,revsymb}
\usepackage{graphicx}
\usepackage{bm}
\usepackage[dvipsnames]{xcolor}
\usepackage{chemformula}
 \usepackage[separate-uncertainty=true,multi-part-units=single,range-phrase={ - }]{siunitx}
\usepackage{bm}
\usepackage{mathtools} 
\usepackage[colorlinks=true,linkcolor=blue,citecolor=blue,urlcolor=blue]{hyperref}
\usepackage{lineno}
\usepackage{soul}

\makeatletter
\def\maketitle{
\@author@finish
\title@column\titleblock@produce
\suppressfloats[t]}
\makeatother

\newcommand\mrm[1]{\mathrm{#1}}

\newcommand{\SLIO}{Sr$_{2-x}$La$_{x}$IrO$_{4}$}
\newcommand{\SLIOptwo}{Sr$_{1.8}$La$_{0.2}$IrO$_{4}$}
\newcommand{\SIO}{Sr$_{2}$IrO$_{4}$}

\newcommand{\EF}{E_{\mrm{F}}}
\newcommand{\kF}{k_{\mrm{F}}}
\newcommand{\vF}{v_{\mrm{F}}}
\newcommand{\GF}{\Gamma_{\mrm{F}}}
\newcommand{\WF}{W_{\mrm{F}}}

\newcommand{\ea}{\textit{et al.}}
\newcommand{\ie}{\textit{i.e.}}

\begin{document}
	
\title{Fermi surface and pseudogap in highly doped Sr$_{2}$IrO$_{4}$}

\author {Y. Alexanian}\email{yann.alexanian@unige.ch}
\affiliation{Department of Quantum Matter Physics, University of Geneva, Geneva, Switzerland}
\author{A. de la Torre}
\affiliation{Department of Physics, Northeastern University, Boston, MA, USA}
\affiliation{Quantum Materials and Sensing Institute, Northeastern University, Burlington, MA USA}
\author{S. McKeown Walker}
\affiliation{Department of Quantum Matter Physics, University of Geneva, Geneva, Switzerland}
\affiliation{Laboratory of Advanced Technology, University of Geneva, Geneva, Switzerland}
\author {M. Straub}
\affiliation{Department of Quantum Matter Physics, University of Geneva, Geneva, Switzerland}
\author {G. Gatti}
\affiliation{Department of Quantum Matter Physics, University of Geneva, Geneva, Switzerland}
\author {A. Hunter}
\affiliation{Department of Quantum Matter Physics, University of Geneva, Geneva, Switzerland}
\author {S. Mandloi}
\affiliation{Department of Quantum Matter Physics, University of Geneva, Geneva, Switzerland}
\author{E. Cappelli}
\affiliation{Department of Quantum Matter Physics, University of Geneva, Geneva, Switzerland}
\author{S. Riccò}
\affiliation{Department of Quantum Matter Physics, University of Geneva, Geneva, Switzerland}
\author{F. Y. Bruno}
\affiliation{GFMC, Departamento de Física de Materiales, Universidad Complutense de Madrid, Madrid, Spain}
\author{M. Radovic}
\affiliation{Swiss Light Source, Paul Scherrer Institut, Villigen, Switzerland}
\author{N. C. Plumb}
\affiliation{Swiss Light Source, Paul Scherrer Institut, Villigen, Switzerland}
\author{M. Shi}
\affiliation{Swiss Light Source, Paul Scherrer Institut, Villigen, Switzerland}
\affiliation{Center for Correlated Matter and School of Physics, Zhejiang University, Hangzhou, China}
\author{J. Osiecki}
\affiliation{MAX IV Laboratory, Lund University, Lund, Sweden}
\author{C. Polley}
\affiliation{MAX IV Laboratory, Lund University, Lund, Sweden}
\author{T. K. Kim}
\affiliation{Diamond Light Source, Harwell Campus, Didcot, UK}
\author{P. Dudin}
\affiliation{Diamond Light Source, Harwell Campus, Didcot, UK}
\affiliation{Synchrotron SOLEIL, Gif sur Yvette, France}
\author{M. Hoesch}
\affiliation{Deutsches Elektronen-Sychrotron DESY, Hamburg, Germany}
\author{R. S. Perry}
\affiliation{ISIS Pulsed Neutron and Muon Source, STFC Rutherford Appleton Laboratory, Harwell Campus, Didcot, UK}
\affiliation{London Centre for Nanotechnology and Department of Physics and Astronomy, University College London, London, UK}
\author {A. Tamai}
\affiliation{Department of Quantum Matter Physics, University of Geneva, Geneva, Switzerland}
\author {F. Baumberger}
\affiliation{Department of Quantum Matter Physics, University of Geneva, Geneva, Switzerland}
\affiliation{Swiss Light Source, Paul Scherrer Institut, Villigen, Switzerland}

	
\begin{abstract}

The fate of the Fermi surface in bulk electron-doped \SIO\ remains elusive, as does the origin and extension of its pseudogap phase. Here, we use high-resolution angle-resolved photoelectron spectroscopy (ARPES) to investigate the electronic structure of \SLIO\ up to $x=0.2$, a factor of two higher than in previous work. 
We find that the antinodal pseudogap persists up to the highest doping level, and thus beyond the sharp increase in Hall carrier density to $\simeq 1+x$ recently observed above $x^{*}\simeq0.16$~\cite{Hsu2024}. This suggests that doped iridates host a unique phase of matter in which a large Hall density coexists with an anisotropic pseudogap, breaking up the Fermi surface into disconnected arcs.
The temperature boundary of the pseudogap is 
$T^{*}\simeq\SI{200}{\kelvin}$ for $x=0.2$, comparable to cuprates 
and to the energy scale of short range antiferromagnetic correlations in cuprates and iridates.

\end{abstract}

\maketitle

\begin{figure*}[htb]
\includegraphics[width=1\textwidth]{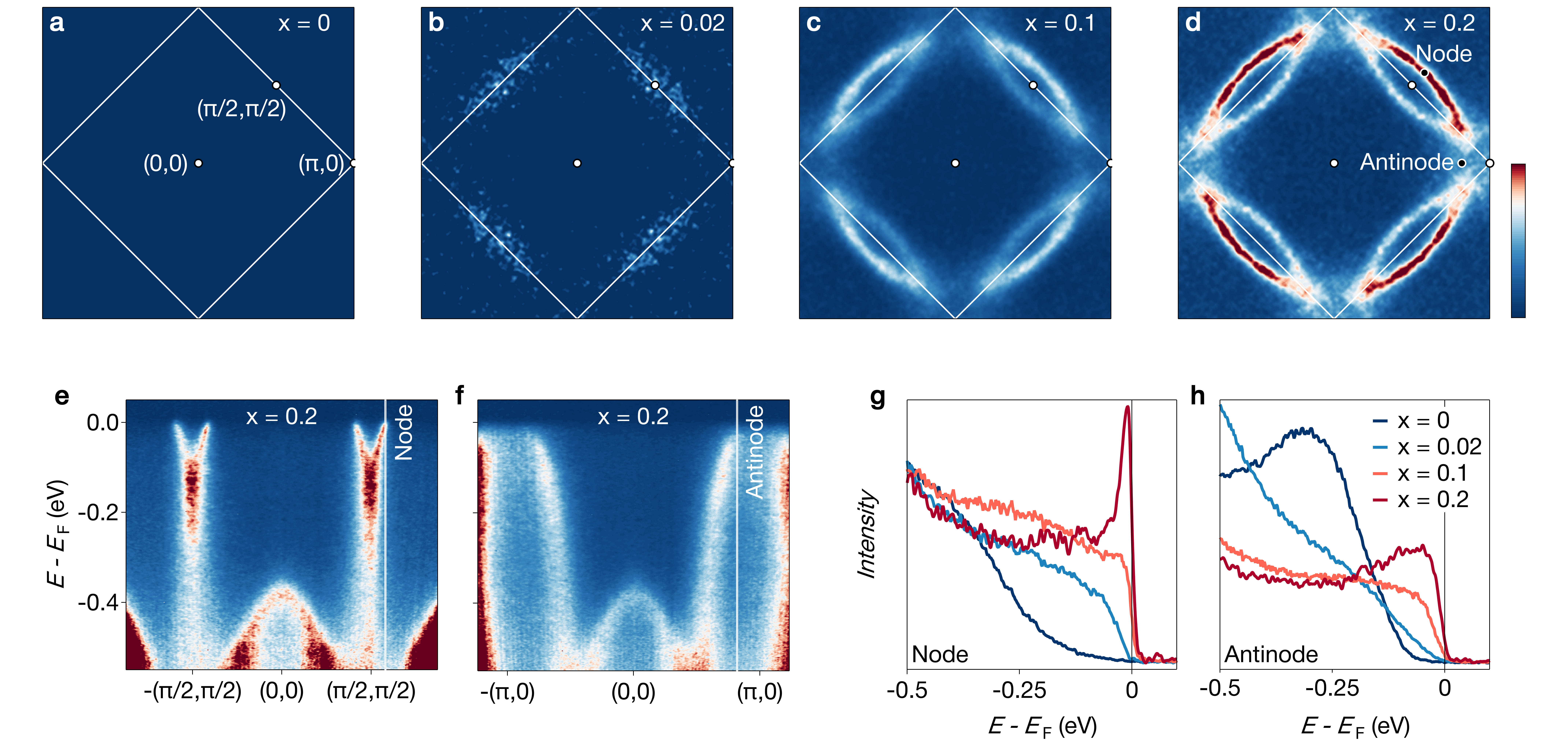}
\caption{\textbf{Collapse of the spin-orbit driven Mott insulating ground state and emergence of the pseudogap state in \SLIO.} \textbf{a-d} Fermi surfaces for $x=0$, $x=0.02$, $x=0.1$, and $x=0.2$, respectively.  
Data were measured at $T\simeq\SI{50}{\kelvin}$ (\textbf{a,b}) and $T\simeq\SI{10}{\kelvin}$ (\textbf{c,d}) with a photon energy $h\nu = \SI{100}{\electronvolt}$ and have been fourfold rotationally averaged. \textbf{e,f} ARPES band dispersion of \SLIOptwo\ along the nodal $(0,0) - (\pi,\pi)$ and antinodal $(0,0) - (\pi,0)$ directions, illustrating the nodal-antinodal dichotomy. 
\textbf{g,h} Doping dependence of energy distribution curves (EDCs) at the nodal and antinodal positions indicated in \textbf{d-f}.}
\label{fig:Fig1}
\end{figure*}



\section{Introduction}

\noindent 
The pseudogap (PG) in hole doped cuprates is one of the most enigmatic properties of correlated electron systems.
A pragmatic definition of a PG, adopted throughout this article, is the existence of a sharp suppression of spectral weight at low energy scales.
ARPES experiments established that the cuprate PG is anisotropic and selectively suppresses the low-energy spectral weight near $(\pi,0)$, leaving apparent Fermi arcs extending out from the node along the Brillouin zone diagonal~\cite{Norman1998,Shen2005}. However, the origin of the PG and its relation with the rich phase diagram of cuprates remain controversial, not least because there is little thermodynamic evidence for a genuine phase transition at the critical doping $p^{*}$ and temperature $T^{*}$ where the PG closes~\cite{Lee2006,Civelli2005,Sordi2007,Proust2019}.
When superconductivity is suppressed in high magnetic fields, several cuprate families show a strong peak in the electronic specific heat near $p^{*}$, in some cases accompanied by a $\log(1/T)$ dependence at low temperatures.
Around the same doping level, the Hall carrier density increases from the doping $p$ to $1+p$~\cite{Badoux2016,Putzke2021}. 
Whether these signatures are caused by a quantum critical point associated with the closure of the PG remains debated.

\vspace{0.2cm}

The recent observation by Hsu~\ea\ of similar anomalies in the electronic specific heat and Hall density in the electron doped iridate \SLIO\ provides complementary insight into these open questions~\cite{Hsu2024}. 
Undoped \SIO\ is a single band antiferromagnetic (AF) insulator, commonly described as a pseudospin $J_{\mrm{eff}} = 1/2$ Mott state \cite{Kim2008,Kim2009,Jin2009}, although other interpretations have been put forward~\cite{Choi2024}.
A minimal model of electron doped \SIO{} remarkably
resembles that of hole doped cuprates~\cite{Wang2011}.
ARPES experiments at low La (\ie{} electron) doping revealed a PG with the same anisotropy in momentum space known from hole doped cuprates~\cite{delaTorre2015,Brouet2015,Peng2022}. Crucially, though, \SLIO\ shows no signs of superconductivity up to the highest doping of $x=0.2$ investigated so far~\cite{Wang2018}. 
%
On the other hand, short range AF spin correlations in \SLIO\ closely reflect spin excitations in cuprates~\cite{Kim2012,Pincini2017,Saylor1989,Hayden1991}, although no charge or spin orders were found in iridates. These observations prompted suggestions that the PG in iridates is driven by magnetic fluctuations~\cite{delaTorre2015,Hsu2024}.

\vspace{0.2cm}

Hsu~\ea\ interpreted the anomalies in specific heat and Hall density as a signature of the closure of the PG at a critical La doping $x^{*} \simeq 0.16$ \cite{Hsu2024}. This is qualitatively consistent with an ARPES study of K surface-doped \SIO\ that found a transition from a pseudogapped regime to a conventional large Fermi surface (FS) around a K coverage of $\sim 0.85$ monolayer (ML)~\cite{Kim2014}. However, it is unclear whether the K/AF-insulator interface is representative of highly bulk doped samples, which are metallic and paramagnetic. In bulk electron doped samples there is thus far no direct evidence for a closure of the PG.


Here, we report high-resolution ARPES data from the same batch of \SLIO\ samples studied by Hsu~\ea~\cite{Hsu2024}. 
We find that the electronic structure evolves smoothly across $x^{*}$. Quasiparticle coherence increases monotonously with doping while the nodal Fermi velocity is largely constant. Most importantly, the PG remains open up to at least $x=0.2$
and thus beyond the Hall density crossover reported by Hsu~\ea~\cite{Hsu2024} at $x^{*} \simeq 0.16$. This demonstrates that a pseudogapped state can coexist with a Hall carrier density of $\simeq 1 + x$, conventionally interpreted as a large closed Fermi surface.
We further show that for $x=0.2$, the PG vanishes around $T^{*}\simeq\SI{200}{\kelvin}$, comparable to $T^{*}$ of cuprates and to the Néel transition temperature of the undoped compounds.

\section{Results}

\subsection{Doping dependence}


Fig.~\ref{fig:Fig1} illustrates the doping dependence of the electronic structure of \SLIO\ from the pristine compound to $x = 0.2$. 
Incoherent spectral weight near the Fermi energy $\EF$ first appears at $x=0.02$. However, a defined PG state with coherent Fermi arcs stretching from the nodal $(0,0) - (\pi,\pi)$ direction and evolving into incoherent antinodal excitations only emerges at $x=0.1$.
The transition from an insulating to a pseudogapped state and the persistence of the nodal-antinodal dichotomy up to $x=0.2$ are evident from the energy distribution curves (EDCs) in Figs.~\hyperref[fig:Fig1]{\ref*{fig:Fig1}g}, \hyperref[fig:Fig1]{\ref*{fig:Fig1}h}. EDCs at the antinode show that although the leading edge of spectral weight shifts closer to $\EF$ with increasing doping, a gap persists across the entire doping range.

In contrast, nodal EDCs display the Fermi-Dirac cutoff of ungapped excitations for $x=0.1$ and $x=0.2$. Notably, the quasiparticle peak -- nearly absent for $x=0.1$ -- becomes well-defined at $x=0.2$. This is a signature of a striking increase of the nodal quasiparticle coherence with doping. 
These ARPES signatures of the PG state of \SIO\ are highly reminiscent of hole doped cuprates. Yet there are two key differences. First, the FS in \SIO\ is electron-like and centered at (0,0), contrasting with the hole-like FS at $(\pi,\pi)$ of the cuprates.
Second, a doubling of the in-plane unit cell arising from a rotation of the oxygen octahedra causes a back folding of the FS in \SLIO~\cite{Crawford1994,Ye2013,delaTorre2015}.

The emergence of a distinct coherent nodal quasiparticle peak from the broad incoherent features dominating the spectra at half-filling is also a key feature of pseudogapped cuprates. There, it appears at doping levels deep into the superconducting state but below the critical doping where the PG terminates~\cite{Shen2004,Fournier2010}. This indicates that our highly doped iridates are approaching the pseudogap doping boundary, and that further doping could lead to the predicted ungapped but still nodal-antinodal differentiated state \cite{Moutenet2018}. 
%
It further highlights the conspicuous lack of superconductivity throughout the iridate pseudogap phase.

\begin{figure}[tbp]
\includegraphics[width=1\columnwidth]{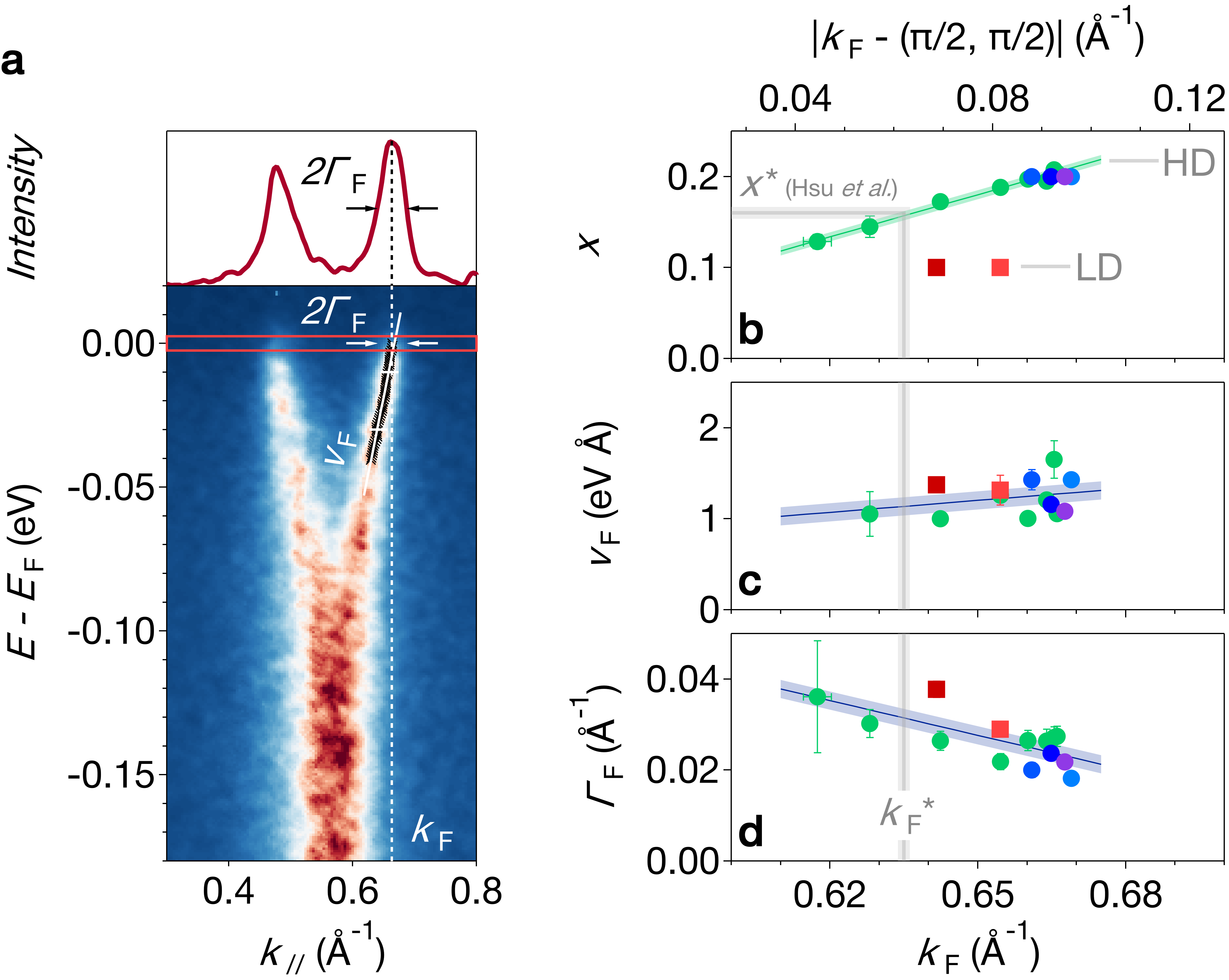 }
\caption{\textbf{Continuous evolution of the electronic structure of \SLIO\ with doping.} \textbf{a} Experimental definition of Fermi momentum $\kF$, Fermi velocity $v_{\mrm{F}}$, and scattering rate at the Fermi energy $\Gamma_{\mrm{F}}$. See methods for details. 
\textbf{b} Lanthanum concentration $x$ versus Fermi momentum $\kF$. 
The green markers represent experimental data measured on a single HD sample which showed a smooth spatial variation of the La doping (see Supplementary Information~A). A linear fit of these data (green line) defines the critical Fermi momentum $k_{\mrm{F}}^{*}$ corresponding to the critical chemical doping $x^{*}$ of Ref. \cite{Hsu2024} (light grey line). $\kF$ measurements on other samples are represented by different colored markers - red squares for lightly doped (LD) and blue/purple dots for highly doped (HD) samples. \textbf{c,d} Fermi velocity $v_{\mrm{F}}$ and scattering rate at the Fermi energy $\Gamma_{\mrm{F}}$ as a function of the Fermi momentum $\kF$. 
Doping values $x$ in (\textbf{b}) have been averaged over the ARPES beam spot diameter of $\SI{10}{\micro\meter}$ and are given with their respective standard deviations, which are often smaller than the marker size. Error bars of $\kF$, $v_{\mrm{F}}$ and $\Gamma_{\mrm{F}}$ in (\textbf{b}-\textbf{d}) represent standard deviations of the fits and are often smaller than the marker size.
}
\label{fig:Fig2_1}
\end{figure}

\begin{figure}[tb]
\includegraphics[width=0.953\columnwidth]{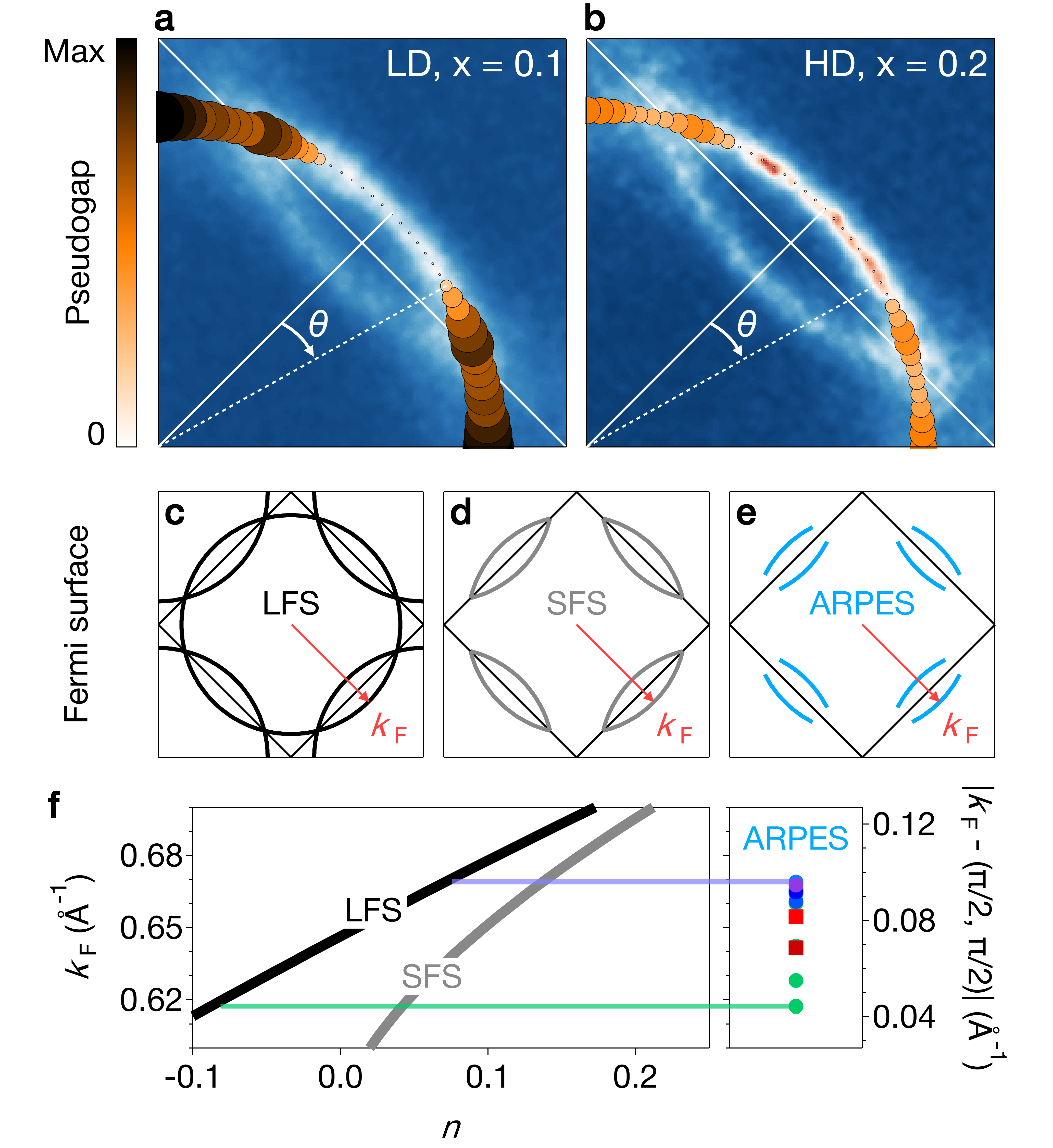 }
\caption{\textbf{Pseudogap state of \SLIO.} \textbf{a,b} Spectral weight suppression in the PG (represented by both marker size and color) overlaid on a quadrant of the Fermi surface for $x=0.1$ and $x=0.2$. A common scale based on the larger value observed for $x=0.1$ is used for both panels. \textbf{c} Schematic representation of the large (LFS) and \textbf{d} small (SFS) Fermi surface scenarios. \textbf{e} Sketch of the actual Fermi surface measured by ARPES. \textbf{f} Left: carrier density per Ir $n$ versus Fermi momentum $\kF$ in the large (black line) and small (grey line) scenarios. Right: range of $\kF$ measured by ARPES in this study for samples with $x$ between $0.1$ and $0.2$, as shown in Figs.~\hyperref[fig:Fig2_1]{\ref*{fig:Fig2_1}b-d}.
}
\label{fig:Fig2_2}
\end{figure}

Fig.~\ref{fig:Fig2_1} quantifies the evolution of the electronic structure of \SLIO\ with doping. To this end, we measured several samples grown with two different procedures (see Methods): lightly doped (nominal doping $x=0.1$) and highly doped (nominal doping $x=0.2$) samples, referred to as LD and HD, respectively. 
Crucially, our HD samples come from the same batch as those measured by Hsu~\ea~\cite{Hsu2024}, allowing for a straightforward comparison of results. 
We determined the precise La content $x$ in both sets with energy-dispersive x-ray spectroscopy (EDX). These measurements revealed a slow variation in $x$ near the edges of some HD samples. 
We exploit this smooth gradient to track the doping dependence of the nodal electronic structure by combining the EDX analysis with spatially resolved $\mu$-spot ARPES measurements (see Supplementary Information~A). 
We first determine the nodal Fermi wave vector $\kF(x)$ (shown as $x(\kF)$ in Fig.~\hyperref[fig:Fig2_1]{\ref*{fig:Fig2_1}b}) from fits to momentum distribution curves (MDCs).
This reveals a clear dichotomy between LD and HD samples, implying that in one or both cases, the itinerant carrier density differs from the La concentration. We will discuss this point in more detail in Fig~\ref{fig:Fig2_2}.
Importantly, though, for the HD samples, also studied by Hsu~\ea~\cite{Hsu2024}, we observe a clean linear evolution of $\kF$ with $x$, without any discontinuity at $x^{*} \simeq 0.16$. 
This allows for a precise determination of the critical nodal $\kF^{*}$ corresponding to the critical doping $x^{*}$ identified in Ref.~\cite{Hsu2024}.

We next determine the nodal Fermi velocity $v_{\mrm{F}}$ and the scattering rate at the Fermi energy $\Gamma_{\mrm{F}}$ from MDC fits (see Methods for details of the analysis).
When plotted against $\kF$, rather than $x$ (Figs.~\hyperref[fig:Fig2_1]{\ref*{fig:Fig2_1}c}, \hyperref[fig:Fig2_1]{\ref*{fig:Fig2_1}d}), $v_{\mrm{F}}$ and $\Gamma_{\mrm{F}}$ of LD and HD samples collapse onto a single curve. This shows that $\kF$ is a more reliable indicator of the electronic state than the La concentration.

As $\kF$ and thus the effective electron doping increases, we only observe a small gradual increase in $\vF$, with no abrupt changes detected at $\kF^{*}$. 
The reduction of the  scattering rate with increasing $\kF$ is more pronounced, in line with the significant rise in nodal quasiparticle coherence observed between the LD, $x=0.1$ and the HD, $x=0.2$ samples (Fig.~\hyperref[fig:Fig1]{\ref*{fig:Fig1}g}). 
Figs.~\hyperref[fig:Fig2_1]{\ref*{fig:Fig2_1}c}, \hyperref[fig:Fig2_1]{\ref*{fig:Fig2_1}d} suggest a continuous evolution of $v_{\mrm{F}}$ and $\Gamma_{\mrm{F}}$ across the critical doping $\kF^{*}$. 
However, we presently have few data points only with $\kF<\kF^{*}$ for which the Fermi velocity and scattering rate remain well defined.
We thus cannot fully exclude a more pronounced change at $\kF^{*}$.

Hsu~\ea\ reported a strong enhancement of the electronic specific heat coefficient $\gamma$ of \SLIO\ over an extended doping range of $x\simeq 0.12 - 0.17$.  
They further point out that their data show no signs of a logarithmic divergence of the specific heat, typical for quantum critical systems~\cite{Hsu2024}.
In a conventional quasi-2D metal, $\gamma\propto m^{*}\propto 1/v_{\mrm{F}}$ where $m^{*}$ is the quasiparticle effective mass. 
An enhancement of $\gamma$ thus corresponds to an enhanced effective mass $m^{*}$ and a correspondingly reduced Fermi velocity $v_{\mrm{F}}$. Our data are difficult to reconcile with such an interpretation of the measured electronic specific heat. 
Specifically, Fig.~\hyperref[fig:Fig2_1]{\ref*{fig:Fig2_1}c} shows that the nodal Fermi velocity remains nearly constant from $x^{*}/ \kF^{*}$
up to the highest doping of $x=0.2$. In contrast, Hsu~\ea\ report a roughly 6-fold decrease of $\gamma$ over the same doping range.
Our direct measurements of the Fermi surface further exclude a Lifshitz transition and associated divergence in the single particle density of states in the relevant doping range. 


In Figs.~\hyperref[fig:Fig2_2]{\ref*{fig:Fig2_2}a}, \hyperref[fig:Fig2_2]{\ref*{fig:Fig2_2}b} we quantify the spectral weight suppression in the PG state along the FS (see Supplementary Information~B for details). This shows an extended ungapped region stretching out from the nodal point up to $\theta\simeq\SI{13.5}{\degree}$ for $x=0.1$ and $\theta\simeq\SI{18}{\degree}$ for $x=0.2$, slightly larger than observed in overdoped Bi$_2$Sr$_2$CaCu$_2$O$_{8+\delta}$~\cite{Lee2007}. Importantly, though, the PG sets on deep into the nodal lens pockets for both $x=0.1$ (LD sample) and $x=0.2$ (HD sample). Hence, no conventional closed Fermi surface emerges up to the highest doping of $x=0.2$.
Instead, we find disconnected Fermi arcs extending out from the node, as illustrated schematically in Fig.~\hyperref[fig:Fig2_2]{\ref*{fig:Fig2_2}e}.

Hsu~\ea\ interpret the Hall carrier density crossover from $n_{\mrm{H}}=1+x$ at high doping to $n_{\mrm{H}} =x$ at low doping as a reconstruction of a conventional large FS (LFS) upon entering the PG phase.
An ungapped LFS -- illustrated in Fig.~\hyperref[fig:Fig2_2]{\ref*{fig:Fig2_2}c} -- has a well defined relation of Fermi wave vector $\kF$ and carrier density per Ir $n$ defined by the Luttinger theorem and shown as a black line in Fig.~\hyperref[fig:Fig2_2]{\ref*{fig:Fig2_2}f}. 
A simplified scenario for the PG phase is a small FS (SFS) consisting only of the lens-like nodal electron pockets while the antinodal hole pockets are gapped (Fig.~ \hyperref[fig:Fig2_2]{\ref*{fig:Fig2_2}d}).

Our data question the applicability of these scenarios. 
Assuming a closed SFS appears arbitrary, given that the PG extends deep into the lens-like contours of the Fermi surface maps (Figs.~\hyperref[fig:Fig2_2]{\ref*{fig:Fig2_2}a}, \hyperref[fig:Fig2_2]{\ref*{fig:Fig2_2}b}).
Moreover, within a LFS scenario -- commonly used in cuprates -- our experimental $\kF$ values for La concentrations $0.1<x<0.2$ translate into carrier densities $-0.09<n<0.07$, nearly symmetric around zero doping.

A discrepancy of itinerant carrier densities and dopant concentration it not unusual. It can arise from co-doping from a slightly off-stoichiometric oxygen content or from a partial localization of doped electrons, as it is observed for instance in SrTiO$_3$ 2D electron gases~\cite{McKeownWalker2015}.
However, obtaining effective hole doping from substituting Sr by La is difficult to rationalize.
Moreover, a LFS scenario places the evidently metallic state of the LD, $x=0.1$ sample (Figs.~\hyperref[fig:Fig1]{\ref*{fig:Fig1}c}, \hyperref[fig:Fig2_2]{\ref*{fig:Fig2_2}a}) at $n\simeq 0$ (half filling), which does not appear plausible.
This suggests that the Luttinger theorem does not apply in the pseudogapped state of \SLIO.

It further highlights that care must be taken when comparing doping values of iridates and cuprates. Applying a LFS scenario, all \SLIO{} samples studied in our work and by Hsu~\ea\cite{Hsu2024} are heavily underdoped in the sense that $n$ is significantly smaller than optimal doping in cuprates $(p\simeq0.15)$.
At the same time, our \SLIO{} samples with $x\simeq 0.2$ are overdoped in the sense that they have a large Hall carrier density $n_{\mrm{H}}\simeq 1+x$~\cite{Hsu2024}, which, in cuprates, is observed only above $p^{*}\simeq0.19$~\cite{Badoux2016,Proust2019}.

\subsection{Temperature dependence}

\begin{figure*}[htb]
\includegraphics[width=1\textwidth]{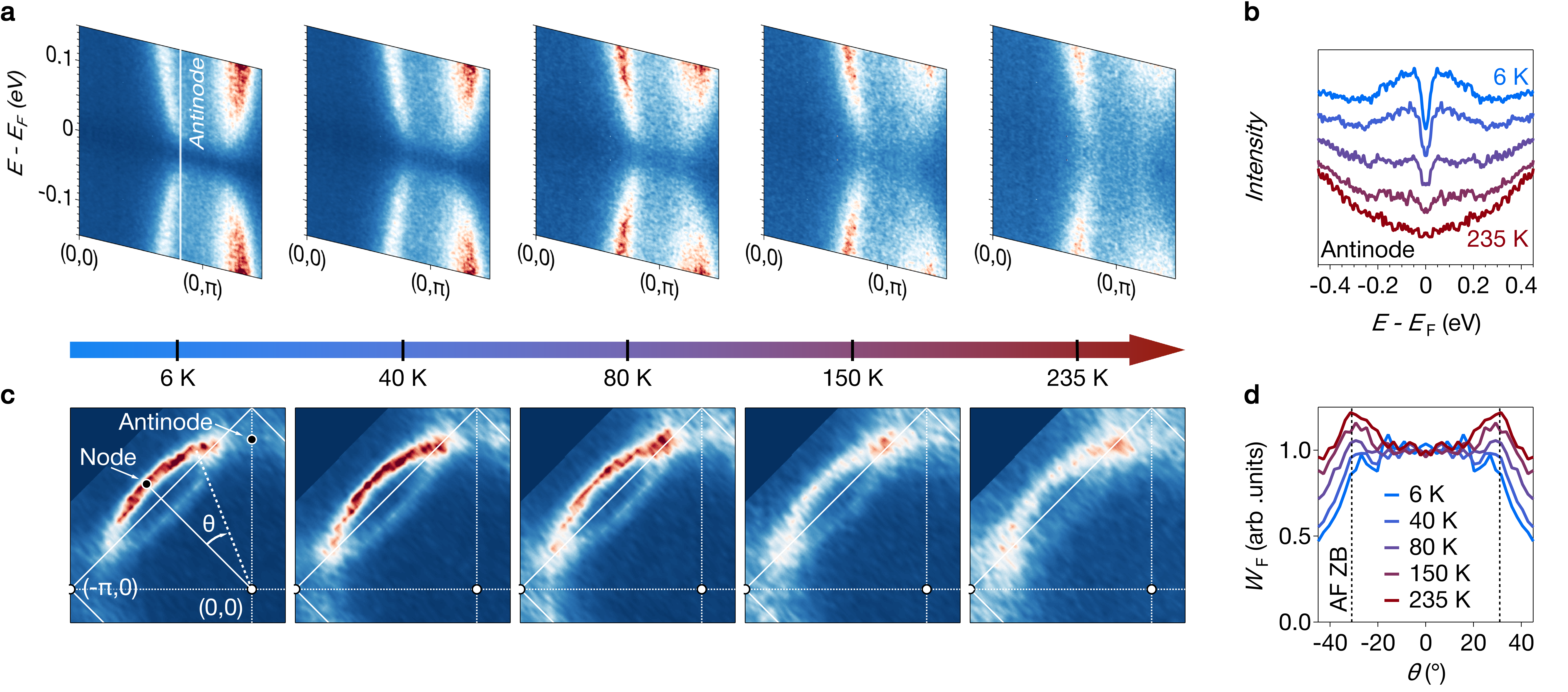 }
\caption{\textbf{Temperature closure of the pseudogap for $x=0.2$.} 
\textbf{a} Temperature dependence of the symmetrized band dispersion in the antinodal direction.  
\textbf{b} EDCs at the antinode (offset for clarity).
\textbf{c} Temperature evolution of the Fermi surface. \textbf{d} 
Temperature dependence of the spectral weight $\WF$ along the Fermi surface.
The angle $\theta$ is defined in panel (\textbf{c}). Data were averaged over $\pm\theta$. See Methods for the procedure used to extract $\WF$. The dashed black lines shows the antiferromagnetic Brillouin zone boundary (AF ZB).}
\label{fig:Fig3}
\end{figure*}

In Fig.~\hyperref[fig:Fig3]{\ref*{fig:Fig3}a} we show the temperature evolution of the antinodal spectral function at the highest doping $x=0.2$ $(\kF \simeq \SI{0.67}{\angstrom^{-1}})$.
The clear gap, persisting all along the antinodal direction at $T=\SI{6}{\kelvin}$, gradually becomes less pronounced as temperature increases and completely disappears by $T=\SI{235}{\kelvin}$. 

The EDCs at the antinodal $\kF$, shown in Fig.~\hyperref[fig:Fig3]{\ref*{fig:Fig3}b} over a large energy range, suggest that the PG disappears primarily by a gradual loss of coherent spectral weight. Similar behavior was reported in cuprates below the critical hole doping $p^{*}$~\cite{Chen2019}. Intriguingly, though, here it is observed at a doping above $x^{*}$ where the Hall carrier density is $\simeq 1+x$~\cite{Hsu2024}.

Measurements along the Brillouin zone diagonal (see Supplementary Information~C) show that the nodal quasiparticle peak is more resilient and persists up to the highest temperature, albeit broadened.
We finally note that low-temperature spectra taken immediately after the temperature-dependent measurements, as shown in Supplementary Information~C, exhibit similar features to those observed before the temperature cycle. This rules out that the PG disappears in the data because of aging effects. 

The vanishing PG further leads to a redistribution of spectral weight $\WF$ along the Fermi surface (see Methods for the determination of $\WF$).
Fig.~\hyperref[fig:Fig3]{\ref*{fig:Fig3}d} shows that the temperature dependence of $\WF$ sets on abruptly at $\theta\simeq 18^{\circ}$ where the ungapped Fermi arc ends. Increasing the angle further, $\WF$ first shows a local maximum around $\theta\simeq\SI{31}{\degree}$ where the Fermi arc and its backfolded replica cross before it decreases towards the antinode.
The suppression of weight towards the antinode is strongest at low temperature and gradually disappears as the highest temperature is approached,
consistent with a vanishing PG. The temperature dependence of the PG further causes a slight shift in angle of the precise position of the local spectral weight maximum.




We remark that the PG reported in Ref.~\cite{Kim2014} for surface K-doped \SIO\ appears to be more fragile than the PG of bulk La-doped \SLIO.
In the K/\SIO\ system the PG was observed to fully close at a K coverage of $\sim 0.85$~ML where $\kF \simeq \SI{0.67}{\angstrom^{-1}}$. 
Our data show that at the same $\kF$, the PG clearly remains open in bulk doped samples. 
Moreover, the temperature boundary of $T^{*}\simeq\SI{200}{\kelvin}$ found here for $x=0.2$ is significantly higher than the maximal $T^{*}=\SIrange{70}{110}{\kelvin}$
at the K/\SIO\ interface observed at a lower doping (coverage of 0.7~ML/$\kF\simeq \SI{0.655}{\angstrom^{-1}}$).
In addition, surface doped \SIO\ shows an apparent $d$-wave gap, suggesting potential high-temperature surface superconductivity~\cite{Kim2015,Yan2015}.
Our data on bulk doped samples exclude a $d$-wave gap of similar magnitude and temperature dependence over the entire doping/$\kF$ range studied in Refs.~\cite{Kim2015,Yan2015}.

\section{Discussion}

The temperature boundary of the PG in bulk electron-doped iridates observed here is strikingly similar to $T^{*}$ in cuprates~\cite{Yoshida2009, CyrChoinire2018}. 
This is intriguing considering the much lower on-site repulsion $U$ and stronger spin-orbit coupling $\lambda_{\textrm{SOC}}$ in iridates~\cite{Moon2009,Kim2012,Wang2013} 
as well as the absence of superconductivity in the samples studied here~\cite{Wang2018,Hsu2024}.
On the other hand, iridates and cuprates show similar energy scales in the magnetic sector.
Magnons in undoped \SIO\ disperse up to $\sim \SI{205}{\meV}$ at $(\pi,0)$~\cite{Kim2012}, comparable to the $\sim \SI{320}{\meV}$ in La$_2$CuO$_4$~\cite{Coldea2001}.
Moreover, spin fluctuations in both iridates~\cite{Gretarsson2016,Pincini2017} and cuprates~\cite{Keimer1992, Drachuck2014} are known to persist up to high doping.
These experimental findings point to an important role of short range antiferromagnetic spin fluctuations in the pseudogap physics of \SLIO.
In turn, this provides further evidence for the importance of AF correlations for the PG in cuprates, which is also seen in numerical solutions of the Hubbard model~\cite{Kyung2006, Gunnarsson2015, Wu2017, Wang2015, Moutenet2018, Simkovic2024}.

We further note a pronounced asymmetry between electron- and hole-doped iridates. A recent study of hole-doped Sr$_{1.93}$K$_{0.07}$IrO$_{4}$ films found a conventional large FS, in sharp contrast to the PG state of \SLIO~\cite{Nelson2020}. This is reminiscent of differences observed in hole- and electron-doped cuprates, although it remains unclear whether the underlying origins are similar. Indeed, both the orbitals involved and the nature of the insulating ground state are different: Cu $e_{g}$ and O $2p$ orbitals with a large charge-transfer gap for the cuprates, spin-orbitally entangled Ir $t_{2g}$ orbitals with small putative Mott-Hubbard gap in the iridate. Consequently, in cuprates doped holes have a different orbital character than doped electrons~\cite{Armitage2010}, while they have different internal degrees of freedom in iridates and thus a different motion in their local magnetic environment~\cite{Parschke2017}. 


In summary, our ARPES measurements of highly doped \SLIO{} samples show a smooth evolution of the electronic structure up to the highest doping of $x\simeq 0.2$. Most importantly, we find that the anisotropic PG of \SLIO{} persists up to $x=0.2$ and thus beyond the crossover to a large Hall carrier density $\simeq 1+x$ reported by Hsu~\ea{} at a critical doping $x^{*}\simeq0.16$.
In cuprates, the crossover of the Hall density from $\simeq x$ to $\simeq 1+x$ occurs at -- or very close to -- the critical doping $p^{*}$ where the PG ends~\cite{Badoux2016,Proust2019}.
Our work, taken together with the results of Hsu~\ea, shows that iridates differ in this crucial aspect of PG phenomenology. At high doping, a large Hall carrier density $\simeq 1+x$ coexists in iridates with an anisotropic pseudogap suppressing the antinodal spectral weight.
Understanding the nature of this unusual state will require further experimental and theoretical work.

%
%

\section{Methods}
\subsection{Crystal growth}

Single crystals with less than $x=0.11$ lanthanum content (LD samples) were grown using a conventional flux cooling method, detailed elsewhere \cite{delaTorre2015}. The lanthanum was measured by energy-dispersive X-ray spectroscopy and electron probe X-ray microanalysis. A flux evaporation method was used for crystals with high lanthanum concentrations (HD samples). For $x=0.2$, SrCl$_{2}$  (Alfa Aesar, anhydrous, 2N$_{5}$) flux was combined with SrCO$_{3}$ (Sigma Aldrich, 4N), IrO$_{2}$ (Alfa Aesar, 3N) and La$_{2}$O$_{3}$ (Sigma Aldrich, 4N) in the ratio $\mrm{IrO}_{2} + 8.5 \mrm{SrCl}_{2} + 1.44 \mrm{SrCO}_{3} + 0.36 \mrm{La}_{2}\mrm{O}_{3}$. The SrCO$_{3}$ was dried at $\SI{600}{\celsius}$ and La$_{2}$O$_{3}$ at $\SI{1000}{\celsius}$ for $\SI{24}{\hour}$. $\SI{0.83}{\gram}$ of IrO$_{2}$ was used for the successful attempts. All materials were ground in an agate mortar and pestle for twenty minutes and loaded into a $\SI{30}{\milli\liter}$ platinum crucible with a loose platinum lid. The crucible was loaded into a standard box furnace at $\SI{700}{\celsius}$, ramped to $\SI{1350}{\celsius}$ in $\SI{3}{\hour}$ and held for $\SI{30}{\hour}$. The furnace was then cooled to $\SI{800}{\celsius}$ at $\SI[per-mode = symbol]{4}{\celsius\per\minute}$, the crucible was removed, and air quenched to room temperature.

The crucible held a mixture of phases, including large ($<\SI{2}{\milli\meter}$), square platelet samples (perhaps twenty per batch) and iridium metal. Little flux remained in the crucible, and the crystals were removed by soaking in warm water. There were often several morphologies, including triangular-platelet and square-platelet. Interestingly, the square-platelet crystals were consistently above $x=0.16$ La content, while the triangular-platelet crystals were less than $x=0.11$. The successful method was highly susceptible to synthesis conditions. Variations in starting mass, time baking, crucible volume and temperature could all affect the outcome, preventing crystal growth of high La concentrations. The baking time needed to be long enough for the majority of the flux to evaporate, driving the growth of large crystals. The temperature window of successful growth was small, perhaps twenty degrees Celsius, vital to achieve high doping. Essentially, the intermediate oxidation state of the iridium required for high lanthanum doping is stabilised by temperature and chemical environment. If the temperature is too low, the iridium will not be sufficiently reduced, and if it is too high, the flux will evaporate too quickly, and the iridium will be reduced to metal. Successful reproduction of our method will require fine-tuning individual set-ups to optimise the temperature and starting mass/growth time.


\subsection{ARPES measurements}

Angle-Resolved Photoemission Spectroscopy (ARPES) experiments were performed at the I05 beamline (Diamond Light Source), the BLOCH beamline (Max IV), and the SIS beamline (Swiss Light Source). Samples were cleaved in ultra-high vacuum conditions ($P<\SI{e-10}{\milli\bar}$) and at low temperatures, $T\simeq\SI{50}{\kelvin}$ for insulating samples ($x=0$, $0.02$), and $T\le\SI{20}{\kelvin}$ for conducting samples ($x\ge0.1$). The energy resolution ranged from $\SI{10}{\milli\electronvolt}$ to $\SI{30}{\milli\electronvolt}$, depending on the specific measurement. The Fermi level was calibrated from reference measurements on a polycrystalline Au sample. \newline
\indent The data shown in Figs.~\ref{fig:Fig1}, \hyperref[fig:Fig2_1]{\ref*{fig:Fig2_1}a}, \hyperref[fig:Fig2_2]{\ref*{fig:Fig2_2}a}, \hyperref[fig:Fig2_2]{\ref*{fig:Fig2_2}b}, and \ref{fig:Fig3} were acquired at the I05 beamline with $\SI{100}{\electronvolt}$ photon energy, linear horizontal (LH) polarization, a vertical analyzer slit, and an ARPES spot size of approximately $\qtyproduct[product-units = power]{50 x 50}{\micro\meter^{2}}$. The light red square and purple dot shown in Figs.~\hyperref[fig:Fig2_1]{\ref*{fig:Fig2_1}b-d} and \hyperref[fig:Fig2_2]{\ref*{fig:Fig2_2}f} were obtained from these data. \newline
\indent The dark red square in Figs.~\hyperref[fig:Fig2_1]{\ref*{fig:Fig2_1}b-d} and \hyperref[fig:Fig2_2]{\ref*{fig:Fig2_2}f} was extracted from data collected at the SIS beamline with $\SI{68}{\electronvolt}$ photon energy, circular right polarization, a horizontal analyzer slit, and a spot size of about $\qtyproduct[product-units = power]{50 x 100}{\micro\meter^{2}}$. \newline
\indent The measurements shown as the blue and green dots in Figs.~\hyperref[fig:Fig2_1]{\ref*{fig:Fig2_1}b-d} and \hyperref[fig:Fig2_2]{\ref*{fig:Fig2_2}f} were performed at the BLOCH beamline, using $\SI{68}{\electronvolt}$ photon energy, LH polarization, a vertical analyzer slit, and a spot size of approximately $\qtyproduct[product-units = power]{10 x 10}{\micro\meter^{2}}$.

\subsection{Details on data analysis} 

The data from undoped and lightly doped samples ($x\le 0.1$) shown in Fig.~\ref{fig:Fig1} are extracted from the dataset of Ref. \cite{delaTorre2015}. All the Fermi surface maps were obtained by integrating the measured intensity in the range $\EF\pm\SI{5}{\milli\electronvolt}$.  \newline 
\indent Fermi momenta $\kF$ and scattering rates at the Fermi energy $\GF$, depicted in Figs.~\hyperref[fig:Fig2_1]{\ref*{fig:Fig2_1}b-d}, were obtained by fitting nodal Momentum Distribution Curves (MDCs) integrated over the range $E=\EF\pm\SI{2.5}{\milli\electronvolt}$. The MDCs, extending over the entire Brillouin zone, were modeled using four Lorentzian peaks with a second-order polynomial background, convolved with a Gaussian profile of full width at half maximum $\Delta E/\vF$ where $\Delta E$ is the energy resolution and $\vF$ the Fermi velocity. The Fermi velocity $\vF$ in Fig.~\hyperref[fig:Fig2_1]{\ref*{fig:Fig2_1}c} was obtained from linear fits of the MDC peak dispersion between $\EF-\SI{30}{\milli\electronvolt}$ and $\EF-\SI{10}{\milli\electronvolt}$. 
Where available, the values of $\kF$, $\GF$, and $\vF$ extracted from both sides of the $\Gamma$ points were averaged. \newline 
\indent To determine the pseudogap area as a function of the angle $\theta$ in Fig.~\hyperref[fig:Fig2_2]{\ref*{fig:Fig2_2}a}, we first removed a second-order polynomial background defined far from $\EF$ (typically from $E-\EF=\SI{-0.3}{\electronvolt}$ to $E-\EF=\SI{-0.05}{\electronvolt}$) to EDCs averaged over the range $\kF\pm\SI{0.015}{\angstrom^{-1}}$ for every Fermi surface angle $\theta$. The background-substracted symmetrized EDCs relative to $\EF$ were fitted with a Gaussian profile, the area of which were averaged for $\pm\theta$ and define the pseudogap spectral weight suppression. \newline 
\indent The spectral weight at the Fermi level $\WF$ shown in Fig.~\hyperref[fig:Fig3]{\ref*{fig:Fig3}d} was obtained by averaging the measured intensity in the range $\EF\pm\SI{10}{\milli\electronvolt}$ and $\kF\pm\SI{0.015}{\angstrom^{-1}}$ at $\pm\theta$ for $\theta\in\left[\SI{-45}{\degree},\SI{45}{\degree}\right]$. For each temperature $T$, this quantity was normalised by the average nodal spectral weight $\WF(\theta\in\left[\SI{0}{\degree},\SI{\pm 5}{\degree}\right],T)$.

\section{Data availability}
The datasets analyzed during this study are available at the Yareta repository of the University of Geneva \cite{data}.

\section{Acknowledgments}
We thank D. van der Marel and K. Wohlfeld for discussion. This work was supported by the Swiss National Science Foundation grants 146995, 165791, 184998. We acknowledge Diamond Light Source for time on Beamline I05 under Proposals SI10348, SI12404, SI17381.
We acknowledge MAX IV Laboratory for time on Beamline Bloch under Proposal No. 20220192 and 20231600. Research conducted at MAX IV, a Swedish national user facility, is supported by the Swedish Research Council under Contract No. 2018-07152, the Swedish Governmental Agency for Innovation Systems under Contract No. 2018-04969, and Formas under Contract No. 2019-0249.
We acknowledge the Paul Scherrer Institut, Villigen, Switzerland for provision of synchrotron radiation beamtime at the SIS beamline of the SLS.

\section{Author contributions}
F.B. and A.T. initiated and coordinated the project. A.d.l.T. and R.S.P. grew the crystals. Y.A., A.d.l.T., S.McK.W., M.S., G.G., A.H., S.M., E.C., S.R., F.Y.B., A.T. and F.B. performed the ARPES measurements with support from M.R., N.C.P., M.S., J.O., C.P., T.K.K., P.D. and M.H. Y.A. analyzed the data and prepared the figures with support from A.T. and F.B. Y.A., A.T. and F.B. wrote the paper with input from all authors.

\section{Competing interests}
The authors declare no competing interests.
%

%


\clearpage

\setcounter{figure}{0} 
\setcounter{equation}{0} 
\setcounter{table}{0}
\setcounter{section}{0} 

\renewcommand{\thefigure}{S\arabic{figure}} 
\renewcommand{\theequation}{S\arabic{equation}} 
\renewcommand{\thetable}{S\arabic{table}}
\renewcommand*{\thesubsection}{\Alph{subsection}}

\makeatletter
\renewcommand{\theHfigure}{supp.\arabic{figure}}
\makeatother

\title{Supplementary Information for `Fermi surface and pseudogap in highly doped Sr$_{2}$IrO$_{4}$'}

\maketitle

\onecolumngrid


\section{A. Sample characterization and doping dependence}
\label{SM_A}

\begin{figure}[h!]
\includegraphics[width=\textwidth]{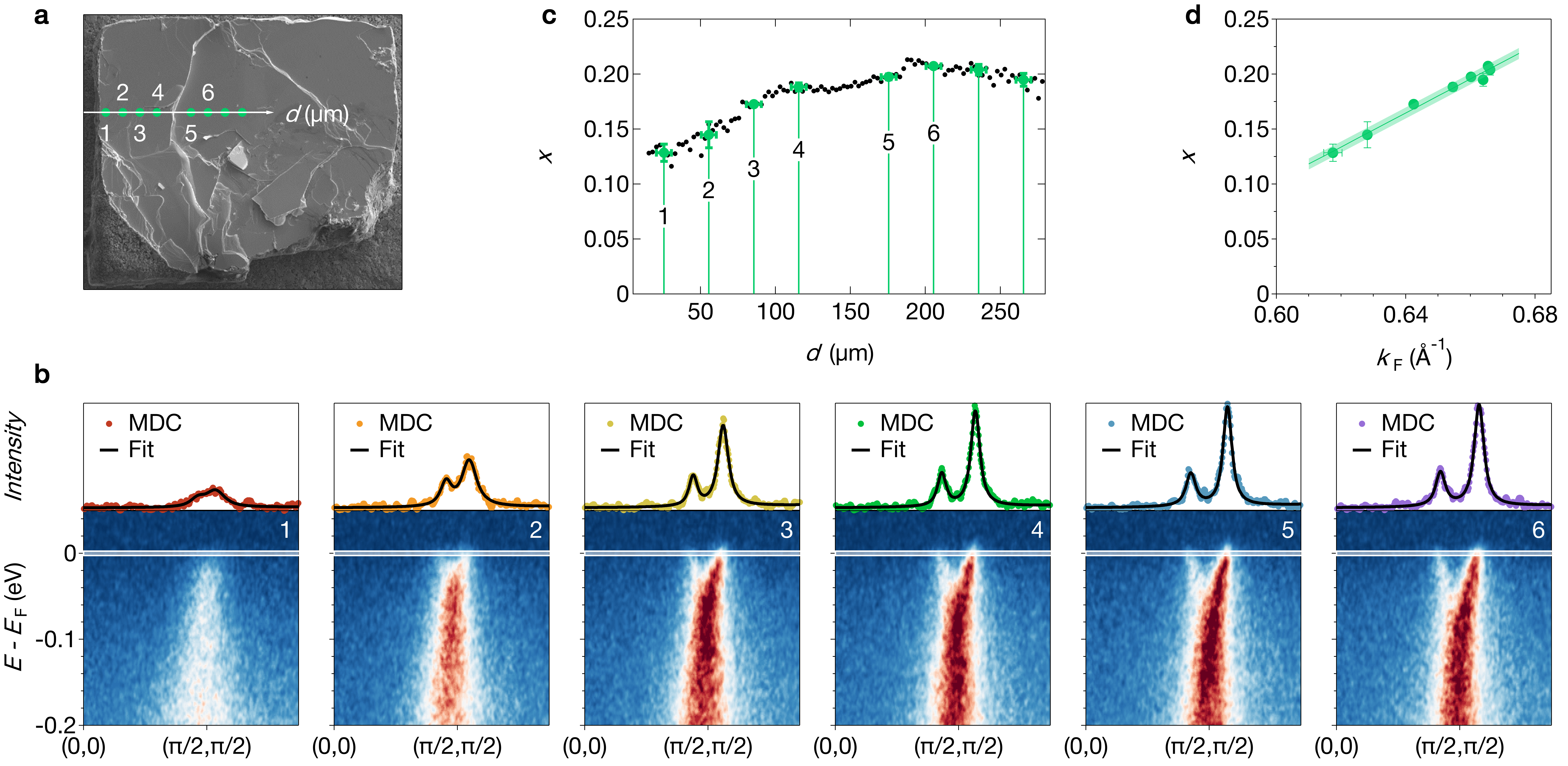}
\caption{\textbf{Spatial dependence of the ARPES spectra measured in an HD sample.} \textbf{a} Scanning electron microscope image of a cleaved sample. The green dots represent the position and size of the ARPES spots at different measurement points. The spot size was approximately $\SI{10}{\micro\meter}$, and the points were spaced by $\SI{30}{\micro\meter}$ along a line (in white). \textbf{b} ARPES cuts and MDCs at the Fermi energy (integrated over the range $E=\EF\pm\SI{2.5}{\milli\electronvolt}$) along the nodal direction measured at different positions on the sample, as shown in panel (\textbf{a}). MDCs are fitted with two Lorentzian peaks and a second-order polynomial background convolved with a Gaussian profile of full width at half maximum determined by the energy resolution. \textbf{c} EDX characterization of the doping value $x$ along the line of the ARPES measurements indicated in panel (\textbf{a}). Green lines indicate positions of the ARPES measurements. Green markers give doping values averaged over the beam spot diameter of $\SI{10}{\micro\meter}$ with standard deviations. \textbf{d} Fermi momentum $\kF$ with standard deviations (often smaller than the marker size) extracted from the MDCs as a function of the chemical doping value $x$ obtained from EDX measurements. The green line is a linear fit to the data.}
\label{fig:SM_Fig1}
\end{figure}

In Fig.~\ref{fig:SM_Fig1}, we present the evolution of the ARPES spectra of \SLIO\ as a function of the ARPES spot position in one of our HD samples, which exhibits inhomogeneous chemical doping $x$. Spectra in the nodal direction measured at several points spaced by $\SI{30}{\micro\meter}$ (see Fig.~\hyperref[fig:SM_Fig1]{\ref*{fig:SM_Fig1}a}) are shown in Fig.~\hyperref[fig:SM_Fig1]{\ref*{fig:SM_Fig1}b}. Significant changes in the electronic structure near the Fermi energy are observed: the two branches around $(\pi/2,\pi/2)$, forming the lenses of the Fermi surface, are indistinguishable near the sample edge but become clearly visible towards its center. This evolution correlates with variations in the lanthanum concentration $x$ along the line, as measured by energy-dispersive X-ray spectroscopy (EDX), and shown in Fig.~\hyperref[fig:SM_Fig1]{\ref*{fig:SM_Fig1}c}. Interestingly, we found that the Fermi momentum $\kF$ (extracted from MDCs at $E=\EF$) increases linearly with the chemical doping value $x$ (Fig.~\hyperref[fig:SM_Fig1]{\ref*{fig:SM_Fig1}d}). Although the chemical doping $x$ cannot be directly related to electron doping (see main text), its evolution in a single type of sample reflects changes in electron doping. Finally, note that this spatial variation of $x$ is typical for our HD samples, whereas LD samples exhibit a more homogeneous doping distribution.


\section{B. Angular dependence of the pseudogap}
\label{SM_B}

\begin{figure}[h!]
\includegraphics[width=\textwidth]{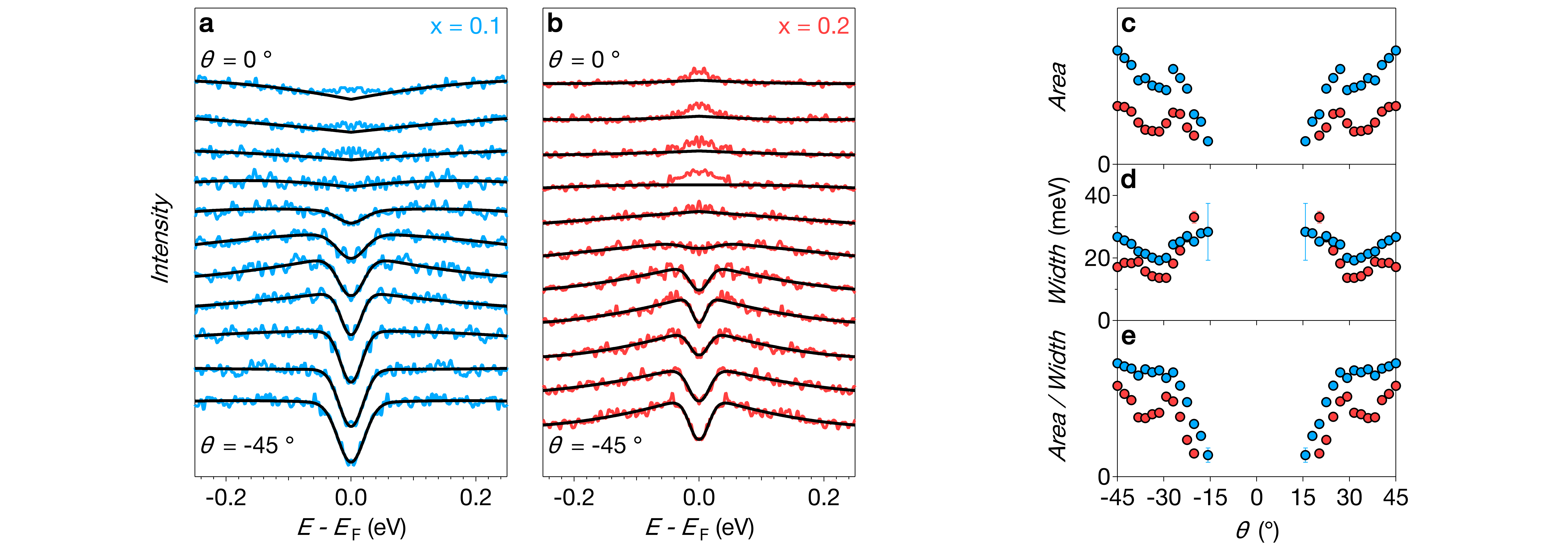}
\caption{\textbf{Angular dependence of the \SLIO\ pseudogap for $x=0.1$ and $x=0.2$.} \textbf{a,b} Fits of the symmetrized EDCs (with respect to $\EF$) measured at $\kF$ with an angle $\theta$ compared to the nodal direction. The fitting procedure is described in the supplementary text. \textbf{c} Pseudogap area $A$, \textbf{d} pseudogap width $\gamma$, and \textbf{e} pseudogap amplitude $A/\gamma$ as a function of the angle $\theta$ compared to the nodal direction. Values obtained for angles $+\theta$ and $-\theta$ were averaged. Error bars represents the standard deviation of the fitted parameters, and are smaller than the marker size except for few data points.}
\label{fig:SM_Fig2}
\end{figure}

In Fig.~\ref{fig:SM_Fig2} we provide additional details on the determination of the angular dependence of the pseudogap shown in Figs.~\hyperref[fig:Fig2_2]{\ref*{fig:Fig2_2}a},~\hyperref[fig:Fig2_2]{\ref*{fig:Fig2_2}b} of the main text. First, we fitted 
a second-order polynomial background $B(E-\EF)$ in the energy interval $\SI{-0.3}{\electronvolt} \le E-\EF \le \SI{-0.05}{\electronvolt}$ of the EDCs measured at $\kF$ for various angles $\theta$ (each EDC corresponds to a color dot in Figs.~\hyperref[fig:Fig2_2]{\ref*{fig:Fig2_2}a},~\hyperref[fig:Fig2_2]{\ref*{fig:Fig2_2}b}). We then removed the background of the EDCs and fitted the symmetrized remaining signal in the energy range $\SI{-0.05}{\electronvolt} \le E-\EF \le \SI{0.05}{\electronvolt}$ with a Gaussian function defined as:
\begin{equation}
    G(E-\EF) = \frac{A}{2\gamma\sqrt{\pi/\ln(16)}}\mathrm{exp}\left(-\left(\frac{E-\EF}{2\gamma}\right)^{2}\ln(16)\right).
\label{Eq:PG}
\end{equation}
Examples of the EDCs and their total fits $B(E-\EF) + G(E-\EF)$ are shown in Fig.~\hyperref[fig:SM_Fig2]{\ref*{fig:SM_Fig2}a} for $x=0.1$ and Fig.~\hyperref[fig:SM_Fig2]{\ref*{fig:SM_Fig2}b} for $x=0.2$. 
The spectral weight suppression in the PG, shown in Figs.~\hyperref[fig:Fig2_2]{\ref*{fig:Fig2_2}a},~\hyperref[fig:Fig2_2]{\ref*{fig:Fig2_2}b} of the main text, is defined as the area $A$.
Figs.~\hyperref[fig:SM_Fig2]{\ref*{fig:SM_Fig2}c-e} show the area $A$ together with the gap width $\gamma$, and the gap amplitude $A/\gamma$.
Interestingly, our fits also indicate that it is the gap amplitude, rather than the gap width, that decreases as the node is approached. Lastly, note the presence of a local maximum (and corresponding local minimum) in the gap area (gap width) around $\SI{\pm30}{\degree}$, where the primitive and backfolded bands cross.

\newpage

\section{C. Complement on the temperature dependence}
\label{SM_C}

\begin{figure}[h!]
\includegraphics[width=1\textwidth]{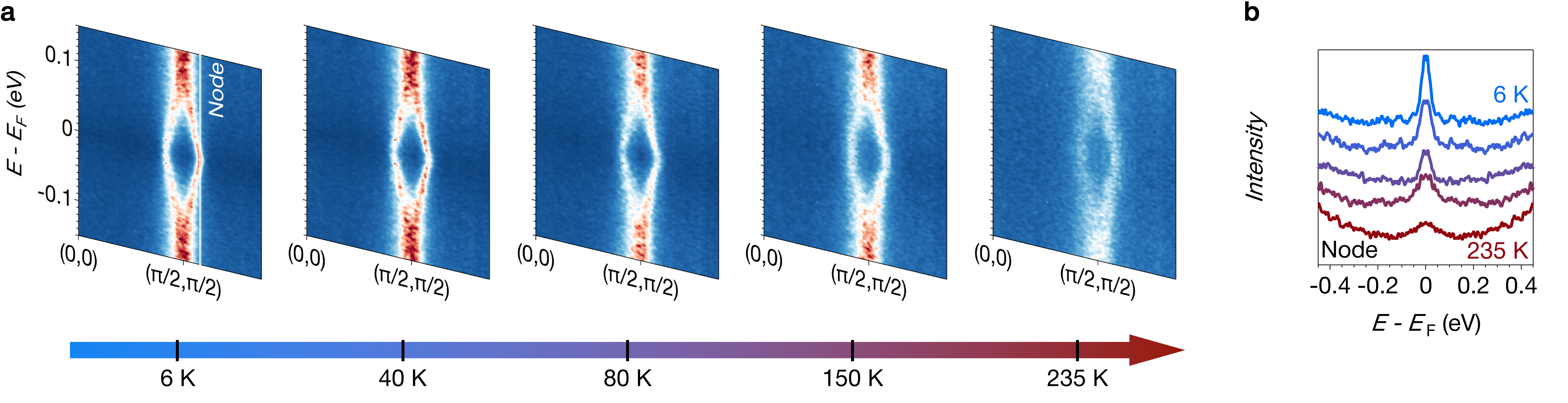}
\caption{\textbf{Temperature dependence of nodal ARPES spectra in \SLIOptwo.} \textbf{a} Temperature dependence of the symmetrized band dispersion in the nodal direction. The node position is indicated by the white line. \textbf{b} EDCs at the node, shifted by a different constant intensity for each temperature.}
\label{fig:SM_Fig3}
\end{figure}

In Fig.~\ref{fig:SM_Fig3} we present temperature-dependent measurements of the electronic structure of \SLIOptwo\ along the nodal direction. Symmetrized cuts from $T=\SI{6}{\kelvin}$ to $T=\SI{235}{\kelvin}$ (see Fig.~\hyperref[fig:SM_Fig3]{\ref*{fig:SM_Fig3}a}) reveal the absence of any gap within this temperature range, in stark contrast to the temperature-dependent pseudogap observed in the antinodal direction (see Fig.~\hyperref[fig:Fig3]{\ref*{fig:Fig3}a} of main text). Moreover, the nodal EDCs at different temperature, shown in Fig.~\hyperref[fig:SM_Fig3]{\ref*{fig:SM_Fig3}b}, demonstrate that the sharp quasiparticle peak visible at $T=\SI{6}{\kelvin}$ broadens with increasing temperature but remains present up to the highest temperature measured. Notably, we observe no significant change when the pseudogap vanishes (between $T=\SI{150}{\kelvin}$ and $T=\SI{235}{\kelvin}$).

\begin{figure}[h!]
\includegraphics[width=1\textwidth]{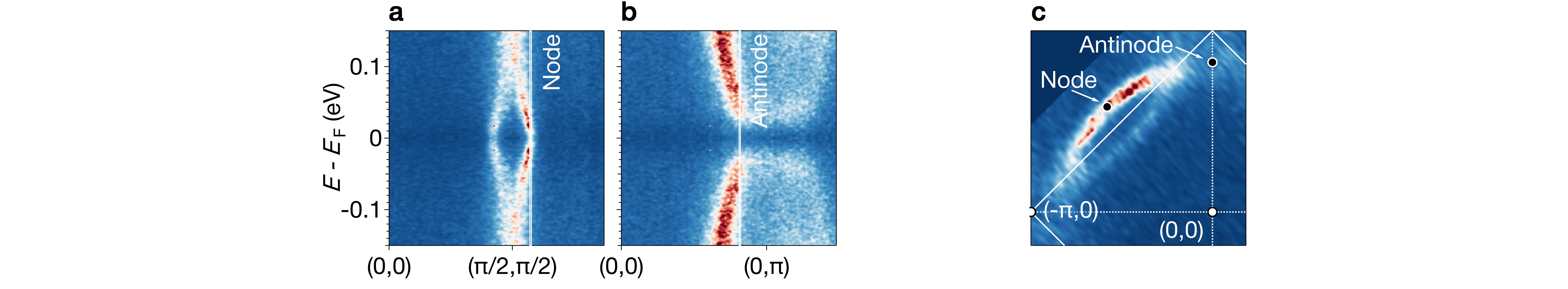}
\caption{\textbf{Low-temperature measurements of \SLIOptwo\ after a complete temperature cycle.} \textbf{a,b} Symmetrized cuts along the nodal and antinodal directions, respectively. \textbf{c} Fermi surface.}
\label{fig:SM_Fig4}
\end{figure}

Low-temperature measurements ($T=\SI{20}{\kelvin}$) taken after the complete temperature cycle from $T=\SI{6}{\kelvin}$ to $T=\SI{235}{\kelvin}$ are shown in Fig.~\ref{fig:SM_Fig4}. Symmetrized cuts along both the nodal and antinodal directions (Figs.~\hyperref[fig:SM_Fig4]{\ref*{fig:SM_Fig4}a},~\hyperref[fig:SM_Fig4]{\ref*{fig:SM_Fig4}b}) as well as the Fermi surface (Fig.~\hyperref[fig:SM_Fig4]{\ref*{fig:SM_Fig4}c}) are similar to those obtained at low temperature before the cycle (see Figs.~\ref{fig:Fig3} and \ref{fig:SM_Fig3}). Most importantly, a pseudogap remains evident at the antinode, while no gap is observed at the node, ruling out any aging effect as an explanation for the pseudogap closure in our data.



\begin{thebibliography}{50}%
\makeatletter
\providecommand \@ifxundefined [1]{%
 \@ifx{#1\undefined}
}%
\providecommand \@ifnum [1]{%
 \ifnum #1\expandafter \@firstoftwo
 \else \expandafter \@secondoftwo
 \fi
}%
\providecommand \@ifx [1]{%
 \ifx #1\expandafter \@firstoftwo
 \else \expandafter \@secondoftwo
 \fi
}%
\providecommand \natexlab [1]{#1}%
\providecommand \enquote  [1]{``#1''}%
\providecommand \bibnamefont  [1]{#1}%
\providecommand \bibfnamefont [1]{#1}%
\providecommand \citenamefont [1]{#1}%
\providecommand \href@noop [0]{\@secondoftwo}%
\providecommand \href [0]{\begingroup \@sanitize@url \@href}%
\providecommand \@href[1]{\@@startlink{#1}\@@href}%
\providecommand \@@href[1]{\endgroup#1\@@endlink}%
\providecommand \@sanitize@url [0]{\catcode `\\12\catcode `\$12\catcode `\&12\catcode `\#12\catcode `\^12\catcode `\_12\catcode `\%12\relax}%
\providecommand \@@startlink[1]{}%
\providecommand \@@endlink[0]{}%
\providecommand \url  [0]{\begingroup\@sanitize@url \@url }%
\providecommand \@url [1]{\endgroup\@href {#1}{\urlprefix }}%
\providecommand \urlprefix  [0]{URL }%
\providecommand \Eprint [0]{\href }%
\providecommand \doibase [0]{https://doi.org/}%
\providecommand \selectlanguage [0]{\@gobble}%
\providecommand \bibinfo  [0]{\@secondoftwo}%
\providecommand \bibfield  [0]{\@secondoftwo}%
\providecommand \translation [1]{[#1]}%
\providecommand \BibitemOpen [0]{}%
\providecommand \bibitemStop [0]{}%
\providecommand \bibitemNoStop [0]{.\EOS\space}%
\providecommand \EOS [0]{\spacefactor3000\relax}%
\providecommand \BibitemShut  [1]{\csname bibitem#1\endcsname}%
\let\auto@bib@innerbib\@empty
\bibitem [{\citenamefont {Hsu}\ \emph {et~al.}(2024)\citenamefont {Hsu}, \citenamefont {Rydh}, \citenamefont {Berben}, \citenamefont {Duffy}, \citenamefont {de~la Torre}, \citenamefont {Perry},\ and\ \citenamefont {Hussey}}]{Hsu2024}%
  \BibitemOpen
  \bibfield  {author} {\bibinfo {author} {\bibfnamefont {Y.-T.}\ \bibnamefont {Hsu}}, \bibinfo {author} {\bibfnamefont {A.}~\bibnamefont {Rydh}}, \bibinfo {author} {\bibfnamefont {M.}~\bibnamefont {Berben}}, \bibinfo {author} {\bibfnamefont {C.}~\bibnamefont {Duffy}}, \bibinfo {author} {\bibfnamefont {A.}~\bibnamefont {de~la Torre}}, \bibinfo {author} {\bibfnamefont {R.~S.}\ \bibnamefont {Perry}},\ and\ \bibinfo {author} {\bibfnamefont {N.~E.}\ \bibnamefont {Hussey}},\ }\bibfield  {title} {\bibinfo {title} {Carrier density crossover and quasiparticle mass enhancement in a doped 5d mott insulator},\ }\href {https://doi.org/10.1038/s41567-024-02564-3} {\bibfield  {journal} {\bibinfo  {journal} {Nature Physics}\ }\textbf {\bibinfo {volume} {20}},\ \bibinfo {pages} {1596} (\bibinfo {year} {2024})}\BibitemShut {NoStop}%
\bibitem [{\citenamefont {Norman}\ \emph {et~al.}(1998)\citenamefont {Norman}, \citenamefont {Ding}, \citenamefont {Randeria}, \citenamefont {Campuzano}, \citenamefont {Yokoya}, \citenamefont {Takeuchi}, \citenamefont {Takahashi}, \citenamefont {Mochiku}, \citenamefont {Kadowaki}, \citenamefont {Guptasarma},\ and\ \citenamefont {Hinks}}]{Norman1998}%
  \BibitemOpen
  \bibfield  {author} {\bibinfo {author} {\bibfnamefont {M.~R.}\ \bibnamefont {Norman}}, \bibinfo {author} {\bibfnamefont {H.}~\bibnamefont {Ding}}, \bibinfo {author} {\bibfnamefont {M.}~\bibnamefont {Randeria}}, \bibinfo {author} {\bibfnamefont {J.~C.}\ \bibnamefont {Campuzano}}, \bibinfo {author} {\bibfnamefont {T.}~\bibnamefont {Yokoya}}, \bibinfo {author} {\bibfnamefont {T.}~\bibnamefont {Takeuchi}}, \bibinfo {author} {\bibfnamefont {T.}~\bibnamefont {Takahashi}}, \bibinfo {author} {\bibfnamefont {T.}~\bibnamefont {Mochiku}}, \bibinfo {author} {\bibfnamefont {K.}~\bibnamefont {Kadowaki}}, \bibinfo {author} {\bibfnamefont {P.}~\bibnamefont {Guptasarma}},\ and\ \bibinfo {author} {\bibfnamefont {D.~G.}\ \bibnamefont {Hinks}},\ }\bibfield  {title} {\bibinfo {title} {Destruction of the fermi surface in underdoped high-tc superconductors},\ }\href {https://doi.org/10.1038/32366} {\bibfield  {journal} {\bibinfo  {journal} {Nature}\ }\textbf {\bibinfo {volume} {392}},\ \bibinfo {pages} {157} (\bibinfo {year}
  {1998})}\BibitemShut {NoStop}%
\bibitem [{\citenamefont {Shen}\ \emph {et~al.}(2005)\citenamefont {Shen}, \citenamefont {Ronning}, \citenamefont {Lu}, \citenamefont {Baumberger}, \citenamefont {Ingle}, \citenamefont {Lee}, \citenamefont {Meevasana}, \citenamefont {Kohsaka}, \citenamefont {Azuma}, \citenamefont {Takano}, \citenamefont {Takagi},\ and\ \citenamefont {Shen}}]{Shen2005}%
  \BibitemOpen
  \bibfield  {author} {\bibinfo {author} {\bibfnamefont {K.~M.}\ \bibnamefont {Shen}}, \bibinfo {author} {\bibfnamefont {F.}~\bibnamefont {Ronning}}, \bibinfo {author} {\bibfnamefont {D.~H.}\ \bibnamefont {Lu}}, \bibinfo {author} {\bibfnamefont {F.}~\bibnamefont {Baumberger}}, \bibinfo {author} {\bibfnamefont {N.~J.~C.}\ \bibnamefont {Ingle}}, \bibinfo {author} {\bibfnamefont {W.~S.}\ \bibnamefont {Lee}}, \bibinfo {author} {\bibfnamefont {W.}~\bibnamefont {Meevasana}}, \bibinfo {author} {\bibfnamefont {Y.}~\bibnamefont {Kohsaka}}, \bibinfo {author} {\bibfnamefont {M.}~\bibnamefont {Azuma}}, \bibinfo {author} {\bibfnamefont {M.}~\bibnamefont {Takano}}, \bibinfo {author} {\bibfnamefont {H.}~\bibnamefont {Takagi}},\ and\ \bibinfo {author} {\bibfnamefont {Z.-X.}\ \bibnamefont {Shen}},\ }\bibfield  {title} {\bibinfo {title} {{Nodal Quasiparticles and Antinodal Charge Ordering in Ca$_{2-x}$Na$_{x}$CuO$_{2}$Cl$_{2}$}},\ }\href {https://doi.org/10.1126/science.1103627} {\bibfield  {journal} {\bibinfo  {journal}
  {Science}\ }\textbf {\bibinfo {volume} {307}},\ \bibinfo {pages} {901} (\bibinfo {year} {2005})}\BibitemShut {NoStop}%
\bibitem [{\citenamefont {Lee}\ \emph {et~al.}(2006)\citenamefont {Lee}, \citenamefont {Nagaosa},\ and\ \citenamefont {Wen}}]{Lee2006}%
  \BibitemOpen
  \bibfield  {author} {\bibinfo {author} {\bibfnamefont {P.~A.}\ \bibnamefont {Lee}}, \bibinfo {author} {\bibfnamefont {N.}~\bibnamefont {Nagaosa}},\ and\ \bibinfo {author} {\bibfnamefont {X.-G.}\ \bibnamefont {Wen}},\ }\bibfield  {title} {\bibinfo {title} {Doping a mott insulator: Physics of high-temperature superconductivity},\ }\href {http://dx.doi.org/10.1103/RevModPhys.78.17} {\bibfield  {journal} {\bibinfo  {journal} {Reviews of Modern Physics}\ }\textbf {\bibinfo {volume} {78}},\ \bibinfo {pages} {17} (\bibinfo {year} {2006})}\BibitemShut {NoStop}%
\bibitem [{\citenamefont {Civelli}\ \emph {et~al.}(2005)\citenamefont {Civelli}, \citenamefont {Capone}, \citenamefont {Kancharla}, \citenamefont {Parcollet},\ and\ \citenamefont {Kotliar}}]{Civelli2005}%
  \BibitemOpen
  \bibfield  {author} {\bibinfo {author} {\bibfnamefont {M.}~\bibnamefont {Civelli}}, \bibinfo {author} {\bibfnamefont {M.}~\bibnamefont {Capone}}, \bibinfo {author} {\bibfnamefont {S.~S.}\ \bibnamefont {Kancharla}}, \bibinfo {author} {\bibfnamefont {O.}~\bibnamefont {Parcollet}},\ and\ \bibinfo {author} {\bibfnamefont {G.}~\bibnamefont {Kotliar}},\ }\bibfield  {title} {\bibinfo {title} {Dynamical breakup of the fermi surface in a doped mott insulator},\ }\href {http://dx.doi.org/10.1103/PhysRevLett.95.106402} {\bibfield  {journal} {\bibinfo  {journal} {Physical Review Letters}\ }\textbf {\bibinfo {volume} {95}} (\bibinfo {year} {2005})}\BibitemShut {NoStop}%
\bibitem [{\citenamefont {Sordi}\ \emph {et~al.}(2007)\citenamefont {Sordi}, \citenamefont {Amaricci},\ and\ \citenamefont {Rozenberg}}]{Sordi2007}%
  \BibitemOpen
  \bibfield  {author} {\bibinfo {author} {\bibfnamefont {G.}~\bibnamefont {Sordi}}, \bibinfo {author} {\bibfnamefont {A.}~\bibnamefont {Amaricci}},\ and\ \bibinfo {author} {\bibfnamefont {M.~J.}\ \bibnamefont {Rozenberg}},\ }\bibfield  {title} {\bibinfo {title} {Metal-insulator transitions in the periodic anderson model},\ }\href {http://dx.doi.org/10.1103/PhysRevLett.99.196403} {\bibfield  {journal} {\bibinfo  {journal} {Physical Review Letters}\ }\textbf {\bibinfo {volume} {99}} (\bibinfo {year} {2007})}\BibitemShut {NoStop}%
\bibitem [{\citenamefont {Proust}\ and\ \citenamefont {Taillefer}(2019)}]{Proust2019}%
  \BibitemOpen
  \bibfield  {author} {\bibinfo {author} {\bibfnamefont {C.}~\bibnamefont {Proust}}\ and\ \bibinfo {author} {\bibfnamefont {L.}~\bibnamefont {Taillefer}},\ }\bibfield  {title} {\bibinfo {title} {The remarkable underlying ground states of cuprate superconductors},\ }\href {https://doi.org/https://doi.org/10.1146/annurev-conmatphys-031218-013210} {\bibfield  {journal} {\bibinfo  {journal} {Annual Review of Condensed Matter Physics}\ }\textbf {\bibinfo {volume} {10}},\ \bibinfo {pages} {409} (\bibinfo {year} {2019})}\BibitemShut {NoStop}%
\bibitem [{\citenamefont {Badoux}\ \emph {et~al.}(2016)\citenamefont {Badoux}, \citenamefont {Tabis}, \citenamefont {Lalibert{\'e}}, \citenamefont {Grissonnanche}, \citenamefont {Vignolle}, \citenamefont {Vignolles}, \citenamefont {B{\'e}ard}, \citenamefont {Bonn}, \citenamefont {Hardy}, \citenamefont {Liang}, \citenamefont {Doiron-Leyraud}, \citenamefont {Taillefer},\ and\ \citenamefont {Proust}}]{Badoux2016}%
  \BibitemOpen
  \bibfield  {author} {\bibinfo {author} {\bibfnamefont {S.}~\bibnamefont {Badoux}}, \bibinfo {author} {\bibfnamefont {W.}~\bibnamefont {Tabis}}, \bibinfo {author} {\bibfnamefont {F.}~\bibnamefont {Lalibert{\'e}}}, \bibinfo {author} {\bibfnamefont {G.}~\bibnamefont {Grissonnanche}}, \bibinfo {author} {\bibfnamefont {B.}~\bibnamefont {Vignolle}}, \bibinfo {author} {\bibfnamefont {D.}~\bibnamefont {Vignolles}}, \bibinfo {author} {\bibfnamefont {J.}~\bibnamefont {B{\'e}ard}}, \bibinfo {author} {\bibfnamefont {D.~A.}\ \bibnamefont {Bonn}}, \bibinfo {author} {\bibfnamefont {W.~N.}\ \bibnamefont {Hardy}}, \bibinfo {author} {\bibfnamefont {R.}~\bibnamefont {Liang}}, \bibinfo {author} {\bibfnamefont {N.}~\bibnamefont {Doiron-Leyraud}}, \bibinfo {author} {\bibfnamefont {L.}~\bibnamefont {Taillefer}},\ and\ \bibinfo {author} {\bibfnamefont {C.}~\bibnamefont {Proust}},\ }\bibfield  {title} {\bibinfo {title} {Change of carrier density at the pseudogap critical point of a cuprate superconductor},\ }\href
  {https://doi.org/10.1038/nature16983} {\bibfield  {journal} {\bibinfo  {journal} {Nature}\ }\textbf {\bibinfo {volume} {531}},\ \bibinfo {pages} {210} (\bibinfo {year} {2016})}\BibitemShut {NoStop}%
\bibitem [{\citenamefont {Putzke}\ \emph {et~al.}(2021)\citenamefont {Putzke}, \citenamefont {Benhabib}, \citenamefont {Tabis}, \citenamefont {Ayres}, \citenamefont {Wang}, \citenamefont {Malone}, \citenamefont {Licciardello}, \citenamefont {Lu}, \citenamefont {Kondo}, \citenamefont {Takeuchi}, \citenamefont {Hussey}, \citenamefont {Cooper},\ and\ \citenamefont {Carrington}}]{Putzke2021}%
  \BibitemOpen
  \bibfield  {author} {\bibinfo {author} {\bibfnamefont {C.}~\bibnamefont {Putzke}}, \bibinfo {author} {\bibfnamefont {S.}~\bibnamefont {Benhabib}}, \bibinfo {author} {\bibfnamefont {W.}~\bibnamefont {Tabis}}, \bibinfo {author} {\bibfnamefont {J.}~\bibnamefont {Ayres}}, \bibinfo {author} {\bibfnamefont {Z.}~\bibnamefont {Wang}}, \bibinfo {author} {\bibfnamefont {L.}~\bibnamefont {Malone}}, \bibinfo {author} {\bibfnamefont {S.}~\bibnamefont {Licciardello}}, \bibinfo {author} {\bibfnamefont {J.}~\bibnamefont {Lu}}, \bibinfo {author} {\bibfnamefont {T.}~\bibnamefont {Kondo}}, \bibinfo {author} {\bibfnamefont {T.}~\bibnamefont {Takeuchi}}, \bibinfo {author} {\bibfnamefont {N.~E.}\ \bibnamefont {Hussey}}, \bibinfo {author} {\bibfnamefont {J.~R.}\ \bibnamefont {Cooper}},\ and\ \bibinfo {author} {\bibfnamefont {A.}~\bibnamefont {Carrington}},\ }\bibfield  {title} {\bibinfo {title} {Reduced hall carrier density in the overdoped strange metal regime of cuprate superconductors},\ }\href
  {https://doi.org/10.1038/s41567-021-01197-0} {\bibfield  {journal} {\bibinfo  {journal} {Nature Physics}\ }\textbf {\bibinfo {volume} {17}},\ \bibinfo {pages} {826} (\bibinfo {year} {2021})}\BibitemShut {NoStop}%
\bibitem [{\citenamefont {Kim}\ \emph {et~al.}(2008)\citenamefont {Kim}, \citenamefont {Jin}, \citenamefont {Moon}, \citenamefont {Kim}, \citenamefont {Park}, \citenamefont {Leem}, \citenamefont {Yu}, \citenamefont {Noh}, \citenamefont {Kim}, \citenamefont {Oh}, \citenamefont {Park}, \citenamefont {Durairaj}, \citenamefont {Cao},\ and\ \citenamefont {Rotenberg}}]{Kim2008}%
  \BibitemOpen
  \bibfield  {author} {\bibinfo {author} {\bibfnamefont {B.~J.}\ \bibnamefont {Kim}}, \bibinfo {author} {\bibfnamefont {H.}~\bibnamefont {Jin}}, \bibinfo {author} {\bibfnamefont {S.~J.}\ \bibnamefont {Moon}}, \bibinfo {author} {\bibfnamefont {J.-Y.}\ \bibnamefont {Kim}}, \bibinfo {author} {\bibfnamefont {B.-G.}\ \bibnamefont {Park}}, \bibinfo {author} {\bibfnamefont {C.~S.}\ \bibnamefont {Leem}}, \bibinfo {author} {\bibfnamefont {J.}~\bibnamefont {Yu}}, \bibinfo {author} {\bibfnamefont {T.~W.}\ \bibnamefont {Noh}}, \bibinfo {author} {\bibfnamefont {C.}~\bibnamefont {Kim}}, \bibinfo {author} {\bibfnamefont {S.-J.}\ \bibnamefont {Oh}}, \bibinfo {author} {\bibfnamefont {J.-H.}\ \bibnamefont {Park}}, \bibinfo {author} {\bibfnamefont {V.}~\bibnamefont {Durairaj}}, \bibinfo {author} {\bibfnamefont {G.}~\bibnamefont {Cao}},\ and\ \bibinfo {author} {\bibfnamefont {E.}~\bibnamefont {Rotenberg}},\ }\bibfield  {title} {\bibinfo {title} {Novel ${J}_{\mathrm{eff}}=1/2$ mott state induced by relativistic spin-orbit
  coupling in {Sr$_{2}$IrO$_{4}$}},\ }\href {http://dx.doi.org/10.1103/PhysRevLett.101.076402} {\bibfield  {journal} {\bibinfo  {journal} {Physical Review Letters}\ }\textbf {\bibinfo {volume} {101}},\ \bibinfo {pages} {076402} (\bibinfo {year} {2008})}\BibitemShut {NoStop}%
\bibitem [{\citenamefont {Kim}\ \emph {et~al.}(2009)\citenamefont {Kim}, \citenamefont {Ohsumi}, \citenamefont {Komesu}, \citenamefont {Sakai}, \citenamefont {Morita}, \citenamefont {Takagi},\ and\ \citenamefont {Arima}}]{Kim2009}%
  \BibitemOpen
  \bibfield  {author} {\bibinfo {author} {\bibfnamefont {B.~J.}\ \bibnamefont {Kim}}, \bibinfo {author} {\bibfnamefont {H.}~\bibnamefont {Ohsumi}}, \bibinfo {author} {\bibfnamefont {T.}~\bibnamefont {Komesu}}, \bibinfo {author} {\bibfnamefont {S.}~\bibnamefont {Sakai}}, \bibinfo {author} {\bibfnamefont {T.}~\bibnamefont {Morita}}, \bibinfo {author} {\bibfnamefont {H.}~\bibnamefont {Takagi}},\ and\ \bibinfo {author} {\bibfnamefont {T.}~\bibnamefont {Arima}},\ }\bibfield  {title} {\bibinfo {title} {{Phase-Sensitive Observation of a Spin-Orbital Mott State in Sr$_{2}$IrO$_{4}$}},\ }\href {http://dx.doi.org/10.1126/science.1167106} {\bibfield  {journal} {\bibinfo  {journal} {Science}\ }\textbf {\bibinfo {volume} {323}},\ \bibinfo {pages} {1329} (\bibinfo {year} {2009})}\BibitemShut {NoStop}%
\bibitem [{\citenamefont {Jin}\ \emph {et~al.}(2009)\citenamefont {Jin}, \citenamefont {Jeong}, \citenamefont {Ozaki},\ and\ \citenamefont {Yu}}]{Jin2009}%
  \BibitemOpen
  \bibfield  {author} {\bibinfo {author} {\bibfnamefont {H.}~\bibnamefont {Jin}}, \bibinfo {author} {\bibfnamefont {H.}~\bibnamefont {Jeong}}, \bibinfo {author} {\bibfnamefont {T.}~\bibnamefont {Ozaki}},\ and\ \bibinfo {author} {\bibfnamefont {J.}~\bibnamefont {Yu}},\ }\bibfield  {title} {\bibinfo {title} {Anisotropic exchange interactions of spin-orbit-integrated states in {Sr$_{2}$IrO$_{4}$}},\ }\href {http://dx.doi.org/10.1103/PhysRevB.80.075112} {\bibfield  {journal} {\bibinfo  {journal} {Physical Review B}\ }\textbf {\bibinfo {volume} {80}},\ \bibinfo {pages} {075112} (\bibinfo {year} {2009})}\BibitemShut {NoStop}%
\bibitem [{\citenamefont {Choi}\ \emph {et~al.}(2024)\citenamefont {Choi}, \citenamefont {Yue}, \citenamefont {Azoury}, \citenamefont {Porter}, \citenamefont {Chen}, \citenamefont {Petocchi}, \citenamefont {Baldini}, \citenamefont {Lv}, \citenamefont {Mogi}, \citenamefont {Su}, \citenamefont {Wilson}, \citenamefont {Eckstein}, \citenamefont {Werner},\ and\ \citenamefont {Gedik}}]{Choi2024}%
  \BibitemOpen
  \bibfield  {author} {\bibinfo {author} {\bibfnamefont {D.}~\bibnamefont {Choi}}, \bibinfo {author} {\bibfnamefont {C.}~\bibnamefont {Yue}}, \bibinfo {author} {\bibfnamefont {D.}~\bibnamefont {Azoury}}, \bibinfo {author} {\bibfnamefont {Z.}~\bibnamefont {Porter}}, \bibinfo {author} {\bibfnamefont {J.}~\bibnamefont {Chen}}, \bibinfo {author} {\bibfnamefont {F.}~\bibnamefont {Petocchi}}, \bibinfo {author} {\bibfnamefont {E.}~\bibnamefont {Baldini}}, \bibinfo {author} {\bibfnamefont {B.}~\bibnamefont {Lv}}, \bibinfo {author} {\bibfnamefont {M.}~\bibnamefont {Mogi}}, \bibinfo {author} {\bibfnamefont {Y.}~\bibnamefont {Su}}, \bibinfo {author} {\bibfnamefont {S.~D.}\ \bibnamefont {Wilson}}, \bibinfo {author} {\bibfnamefont {M.}~\bibnamefont {Eckstein}}, \bibinfo {author} {\bibfnamefont {P.}~\bibnamefont {Werner}},\ and\ \bibinfo {author} {\bibfnamefont {N.}~\bibnamefont {Gedik}},\ }\bibfield  {title} {\bibinfo {title} {{Light-induced insulator-metal transition in Sr$_2$IrO$_4$ reveals the nature of the insulating
  ground state}},\ }\href {https://doi.org/10.1073/pnas.2323013121} {\bibfield  {journal} {\bibinfo  {journal} {Proceedings of the National Academy of Sciences}\ }\textbf {\bibinfo {volume} {121}},\ \bibinfo {pages} {e2323013121} (\bibinfo {year} {2024})}\BibitemShut {NoStop}%
\bibitem [{\citenamefont {Wang}\ and\ \citenamefont {Senthil}(2011)}]{Wang2011}%
  \BibitemOpen
  \bibfield  {author} {\bibinfo {author} {\bibfnamefont {F.}~\bibnamefont {Wang}}\ and\ \bibinfo {author} {\bibfnamefont {T.}~\bibnamefont {Senthil}},\ }\bibfield  {title} {\bibinfo {title} {Twisted hubbard model for {Sr$_{2}$IrO$_{4}$}: Magnetism and possible high temperature superconductivity},\ }\href {http://dx.doi.org/10.1103/PhysRevLett.106.136402} {\bibfield  {journal} {\bibinfo  {journal} {Physical Review Letters}\ }\textbf {\bibinfo {volume} {106}},\ \bibinfo {pages} {136402} (\bibinfo {year} {2011})}\BibitemShut {NoStop}%
\bibitem [{\citenamefont {de~la Torre}\ \emph {et~al.}(2015)\citenamefont {de~la Torre}, \citenamefont {McKeown~Walker}, \citenamefont {Bruno}, \citenamefont {Ricc{\'o}}, \citenamefont {Wang}, \citenamefont {Gutierrez~Lezama}, \citenamefont {Scheerer}, \citenamefont {Giriat}, \citenamefont {Jaccard}, \citenamefont {Berthod}, \citenamefont {Kim}, \citenamefont {Hoesch}, \citenamefont {Hunter}, \citenamefont {Perry}, \citenamefont {Tamai},\ and\ \citenamefont {Baumberger}}]{delaTorre2015}%
  \BibitemOpen
  \bibfield  {author} {\bibinfo {author} {\bibfnamefont {A.}~\bibnamefont {de~la Torre}}, \bibinfo {author} {\bibfnamefont {S.}~\bibnamefont {McKeown~Walker}}, \bibinfo {author} {\bibfnamefont {F.}~\bibnamefont {Bruno}}, \bibinfo {author} {\bibfnamefont {S.}~\bibnamefont {Ricc{\'o}}}, \bibinfo {author} {\bibfnamefont {Z.}~\bibnamefont {Wang}}, \bibinfo {author} {\bibfnamefont {I.}~\bibnamefont {Gutierrez~Lezama}}, \bibinfo {author} {\bibfnamefont {G.}~\bibnamefont {Scheerer}}, \bibinfo {author} {\bibfnamefont {G.}~\bibnamefont {Giriat}}, \bibinfo {author} {\bibfnamefont {D.}~\bibnamefont {Jaccard}}, \bibinfo {author} {\bibfnamefont {C.}~\bibnamefont {Berthod}}, \bibinfo {author} {\bibfnamefont {T.}~\bibnamefont {Kim}}, \bibinfo {author} {\bibfnamefont {M.}~\bibnamefont {Hoesch}}, \bibinfo {author} {\bibfnamefont {E.}~\bibnamefont {Hunter}}, \bibinfo {author} {\bibfnamefont {R.}~\bibnamefont {Perry}}, \bibinfo {author} {\bibfnamefont {A.}~\bibnamefont {Tamai}},\ and\ \bibinfo {author} {\bibfnamefont
  {F.}~\bibnamefont {Baumberger}},\ }\bibfield  {title} {\bibinfo {title} {{Collapse of the Mott Gap and Emergence of a Nodal Liquid in Lightly Doped Sr$_{2}$IrO$_{4}$}},\ }\href {http://dx.doi.org/10.1103/PhysRevLett.115.176402} {\bibfield  {journal} {\bibinfo  {journal} {Physical Review Letters}\ }\textbf {\bibinfo {volume} {115}},\ \bibinfo {pages} {176402} (\bibinfo {year} {2015})}\BibitemShut {NoStop}%
\bibitem [{\citenamefont {Brouet}\ \emph {et~al.}(2015)\citenamefont {Brouet}, \citenamefont {Mansart}, \citenamefont {Perfetti}, \citenamefont {Piovera}, \citenamefont {Vobornik}, \citenamefont {Le~F{\`e}vre}, \citenamefont {Bertran}, \citenamefont {Riggs}, \citenamefont {Shapiro}, \citenamefont {Giraldo-Gallo},\ and\ \citenamefont {Fisher}}]{Brouet2015}%
  \BibitemOpen
  \bibfield  {author} {\bibinfo {author} {\bibfnamefont {V.}~\bibnamefont {Brouet}}, \bibinfo {author} {\bibfnamefont {J.}~\bibnamefont {Mansart}}, \bibinfo {author} {\bibfnamefont {L.}~\bibnamefont {Perfetti}}, \bibinfo {author} {\bibfnamefont {C.}~\bibnamefont {Piovera}}, \bibinfo {author} {\bibfnamefont {I.}~\bibnamefont {Vobornik}}, \bibinfo {author} {\bibfnamefont {P.}~\bibnamefont {Le~F{\`e}vre}}, \bibinfo {author} {\bibfnamefont {F.}~\bibnamefont {Bertran}}, \bibinfo {author} {\bibfnamefont {S.~C.}\ \bibnamefont {Riggs}}, \bibinfo {author} {\bibfnamefont {M.~C.}\ \bibnamefont {Shapiro}}, \bibinfo {author} {\bibfnamefont {P.}~\bibnamefont {Giraldo-Gallo}},\ and\ \bibinfo {author} {\bibfnamefont {I.~R.}\ \bibnamefont {Fisher}},\ }\bibfield  {title} {\bibinfo {title} {Transfer of spectral weight across the gap of {Sr$_{2}$IrO$_{4}$} induced by la doping},\ }\href {http://dx.doi.org/10.1103/PhysRevB.92.081117} {\bibfield  {journal} {\bibinfo  {journal} {Physical Review B}\ }\textbf {\bibinfo {volume}
  {92}},\ \bibinfo {pages} {081117(R)} (\bibinfo {year} {2015})}\BibitemShut {NoStop}%
\bibitem [{\citenamefont {Peng}\ \emph {et~al.}(2022)\citenamefont {Peng}, \citenamefont {Lane}, \citenamefont {Hu}, \citenamefont {Guo}, \citenamefont {Chen}, \citenamefont {Sun}, \citenamefont {Hashimoto}, \citenamefont {Lu}, \citenamefont {Shen}, \citenamefont {Wu}, \citenamefont {Chen}, \citenamefont {Markiewicz}, \citenamefont {Wang}, \citenamefont {Bansil}, \citenamefont {Wilson},\ and\ \citenamefont {He}}]{Peng2022}%
  \BibitemOpen
  \bibfield  {author} {\bibinfo {author} {\bibfnamefont {S.}~\bibnamefont {Peng}}, \bibinfo {author} {\bibfnamefont {C.}~\bibnamefont {Lane}}, \bibinfo {author} {\bibfnamefont {Y.}~\bibnamefont {Hu}}, \bibinfo {author} {\bibfnamefont {M.}~\bibnamefont {Guo}}, \bibinfo {author} {\bibfnamefont {X.}~\bibnamefont {Chen}}, \bibinfo {author} {\bibfnamefont {Z.}~\bibnamefont {Sun}}, \bibinfo {author} {\bibfnamefont {M.}~\bibnamefont {Hashimoto}}, \bibinfo {author} {\bibfnamefont {D.}~\bibnamefont {Lu}}, \bibinfo {author} {\bibfnamefont {Z.-X.}\ \bibnamefont {Shen}}, \bibinfo {author} {\bibfnamefont {T.}~\bibnamefont {Wu}}, \bibinfo {author} {\bibfnamefont {X.}~\bibnamefont {Chen}}, \bibinfo {author} {\bibfnamefont {R.~S.}\ \bibnamefont {Markiewicz}}, \bibinfo {author} {\bibfnamefont {Y.}~\bibnamefont {Wang}}, \bibinfo {author} {\bibfnamefont {A.}~\bibnamefont {Bansil}}, \bibinfo {author} {\bibfnamefont {S.~D.}\ \bibnamefont {Wilson}},\ and\ \bibinfo {author} {\bibfnamefont {J.}~\bibnamefont {He}},\ }\bibfield
  {title} {\bibinfo {title} {Electronic nature of the pseudogap in electron-doped {Sr$_{2}$IrO$_{4}$}},\ }\href {https://doi.org/10.1038/s41535-022-00467-1} {\bibfield  {journal} {\bibinfo  {journal} {npj Quantum Materials}\ }\textbf {\bibinfo {volume} {7}},\ \bibinfo {pages} {58} (\bibinfo {year} {2022})}\BibitemShut {NoStop}%
\bibitem [{\citenamefont {Wang}\ \emph {et~al.}(2018)\citenamefont {Wang}, \citenamefont {Bachar}, \citenamefont {Teyssier}, \citenamefont {Luo}, \citenamefont {Rischau}, \citenamefont {Scheerer}, \citenamefont {de~la Torre}, \citenamefont {Perry}, \citenamefont {Baumberger},\ and\ \citenamefont {van~der Marel}}]{Wang2018}%
  \BibitemOpen
  \bibfield  {author} {\bibinfo {author} {\bibfnamefont {K.}~\bibnamefont {Wang}}, \bibinfo {author} {\bibfnamefont {N.}~\bibnamefont {Bachar}}, \bibinfo {author} {\bibfnamefont {J.}~\bibnamefont {Teyssier}}, \bibinfo {author} {\bibfnamefont {W.}~\bibnamefont {Luo}}, \bibinfo {author} {\bibfnamefont {C.~W.}\ \bibnamefont {Rischau}}, \bibinfo {author} {\bibfnamefont {G.}~\bibnamefont {Scheerer}}, \bibinfo {author} {\bibfnamefont {A.}~\bibnamefont {de~la Torre}}, \bibinfo {author} {\bibfnamefont {R.~S.}\ \bibnamefont {Perry}}, \bibinfo {author} {\bibfnamefont {F.}~\bibnamefont {Baumberger}},\ and\ \bibinfo {author} {\bibfnamefont {D.}~\bibnamefont {van~der Marel}},\ }\bibfield  {title} {\bibinfo {title} {Mott transition and collective charge pinning in electron doped {Sr$_{2}$IrO$_{4}$}},\ }\href {http://dx.doi.org/10.1103/PhysRevB.98.045107} {\bibfield  {journal} {\bibinfo  {journal} {Physical Review B}\ }\textbf {\bibinfo {volume} {98}},\ \bibinfo {pages} {045107} (\bibinfo {year} {2018})}\BibitemShut
  {NoStop}%
\bibitem [{\citenamefont {Kim}\ \emph {et~al.}(2012)\citenamefont {Kim}, \citenamefont {Casa}, \citenamefont {Upton}, \citenamefont {Gog}, \citenamefont {Kim}, \citenamefont {Mitchell}, \citenamefont {van Veenendaal}, \citenamefont {Daghofer}, \citenamefont {van~den Brink}, \citenamefont {Khaliullin},\ and\ \citenamefont {Kim}}]{Kim2012}%
  \BibitemOpen
  \bibfield  {author} {\bibinfo {author} {\bibfnamefont {J.}~\bibnamefont {Kim}}, \bibinfo {author} {\bibfnamefont {D.}~\bibnamefont {Casa}}, \bibinfo {author} {\bibfnamefont {M.~H.}\ \bibnamefont {Upton}}, \bibinfo {author} {\bibfnamefont {T.}~\bibnamefont {Gog}}, \bibinfo {author} {\bibfnamefont {Y.-J.}\ \bibnamefont {Kim}}, \bibinfo {author} {\bibfnamefont {J.~F.}\ \bibnamefont {Mitchell}}, \bibinfo {author} {\bibfnamefont {M.}~\bibnamefont {van Veenendaal}}, \bibinfo {author} {\bibfnamefont {M.}~\bibnamefont {Daghofer}}, \bibinfo {author} {\bibfnamefont {J.}~\bibnamefont {van~den Brink}}, \bibinfo {author} {\bibfnamefont {G.}~\bibnamefont {Khaliullin}},\ and\ \bibinfo {author} {\bibfnamefont {B.~J.}\ \bibnamefont {Kim}},\ }\bibfield  {title} {\bibinfo {title} {Magnetic excitation spectra of {Sr$_{2}$IrO$_{4}$} probed by resonant inelastic x-ray scattering: Establishing links to cuprate superconductors},\ }\href {https://doi.org/10.1103/physrevlett.108.177003} {\bibfield  {journal} {\bibinfo  {journal}
  {Physical Review Letters}\ }\textbf {\bibinfo {volume} {108}},\ \bibinfo {pages} {177003} (\bibinfo {year} {2012})}\BibitemShut {NoStop}%
\bibitem [{\citenamefont {Pincini}\ \emph {et~al.}(2017)\citenamefont {Pincini}, \citenamefont {Vale}, \citenamefont {Donnerer}, \citenamefont {de~la Torre}, \citenamefont {Hunter}, \citenamefont {Perry}, \citenamefont {Moretti~Sala}, \citenamefont {Baumberger},\ and\ \citenamefont {McMorrow}}]{Pincini2017}%
  \BibitemOpen
  \bibfield  {author} {\bibinfo {author} {\bibfnamefont {D.}~\bibnamefont {Pincini}}, \bibinfo {author} {\bibfnamefont {J.~G.}\ \bibnamefont {Vale}}, \bibinfo {author} {\bibfnamefont {C.}~\bibnamefont {Donnerer}}, \bibinfo {author} {\bibfnamefont {A.}~\bibnamefont {de~la Torre}}, \bibinfo {author} {\bibfnamefont {E.~C.}\ \bibnamefont {Hunter}}, \bibinfo {author} {\bibfnamefont {R.}~\bibnamefont {Perry}}, \bibinfo {author} {\bibfnamefont {M.}~\bibnamefont {Moretti~Sala}}, \bibinfo {author} {\bibfnamefont {F.}~\bibnamefont {Baumberger}},\ and\ \bibinfo {author} {\bibfnamefont {D.~F.}\ \bibnamefont {McMorrow}},\ }\bibfield  {title} {\bibinfo {title} {Anisotropic exchange and spin-wave damping in pure and electron-doped {Sr$_{2}$IrO$_{4}$}},\ }\href {http://dx.doi.org/10.1103/PhysRevB.96.075162} {\bibfield  {journal} {\bibinfo  {journal} {Physical Review B}\ }\textbf {\bibinfo {volume} {96}},\ \bibinfo {pages} {075162} (\bibinfo {year} {2017})}\BibitemShut {NoStop}%
\bibitem [{\citenamefont {Saylor}\ \emph {et~al.}(1989)\citenamefont {Saylor}, \citenamefont {Takacs}, \citenamefont {Hohenemser}, \citenamefont {Budnick},\ and\ \citenamefont {Chamberland}}]{Saylor1989}%
  \BibitemOpen
  \bibfield  {author} {\bibinfo {author} {\bibfnamefont {J.}~\bibnamefont {Saylor}}, \bibinfo {author} {\bibfnamefont {L.}~\bibnamefont {Takacs}}, \bibinfo {author} {\bibfnamefont {C.}~\bibnamefont {Hohenemser}}, \bibinfo {author} {\bibfnamefont {J.~I.}\ \bibnamefont {Budnick}},\ and\ \bibinfo {author} {\bibfnamefont {B.}~\bibnamefont {Chamberland}},\ }\bibfield  {title} {\bibinfo {title} {N{\'e}el temperature of stoichiometric {La$_{2}$CuO$_{4}$}},\ }\href {http://dx.doi.org/10.1103/PhysRevB.40.6854} {\bibfield  {journal} {\bibinfo  {journal} {Physical Review B}\ }\textbf {\bibinfo {volume} {40}},\ \bibinfo {pages} {6854} (\bibinfo {year} {1989})}\BibitemShut {NoStop}%
\bibitem [{\citenamefont {Hayden}\ \emph {et~al.}(1991)\citenamefont {Hayden}, \citenamefont {Aeppli}, \citenamefont {Osborn}, \citenamefont {Taylor}, \citenamefont {Perring}, \citenamefont {Cheong},\ and\ \citenamefont {Fisk}}]{Hayden1991}%
  \BibitemOpen
  \bibfield  {author} {\bibinfo {author} {\bibfnamefont {S.~M.}\ \bibnamefont {Hayden}}, \bibinfo {author} {\bibfnamefont {G.}~\bibnamefont {Aeppli}}, \bibinfo {author} {\bibfnamefont {R.}~\bibnamefont {Osborn}}, \bibinfo {author} {\bibfnamefont {A.~D.}\ \bibnamefont {Taylor}}, \bibinfo {author} {\bibfnamefont {T.~G.}\ \bibnamefont {Perring}}, \bibinfo {author} {\bibfnamefont {S.-W.}\ \bibnamefont {Cheong}},\ and\ \bibinfo {author} {\bibfnamefont {Z.}~\bibnamefont {Fisk}},\ }\bibfield  {title} {\bibinfo {title} {High-energy spin waves in {La$_{2}$CuO$_{4}$}},\ }\href {http://dx.doi.org/10.1103/PhysRevLett.67.3622} {\bibfield  {journal} {\bibinfo  {journal} {Physical Review Letters}\ }\textbf {\bibinfo {volume} {67}},\ \bibinfo {pages} {3622} (\bibinfo {year} {1991})}\BibitemShut {NoStop}%
\bibitem [{\citenamefont {Kim}\ \emph {et~al.}(2014)\citenamefont {Kim}, \citenamefont {Krupin}, \citenamefont {Denlinger}, \citenamefont {Bostwick}, \citenamefont {Rotenberg}, \citenamefont {Zhao}, \citenamefont {Mitchell}, \citenamefont {Allen},\ and\ \citenamefont {Kim}}]{Kim2014}%
  \BibitemOpen
  \bibfield  {author} {\bibinfo {author} {\bibfnamefont {Y.~K.}\ \bibnamefont {Kim}}, \bibinfo {author} {\bibfnamefont {O.}~\bibnamefont {Krupin}}, \bibinfo {author} {\bibfnamefont {J.~D.}\ \bibnamefont {Denlinger}}, \bibinfo {author} {\bibfnamefont {A.}~\bibnamefont {Bostwick}}, \bibinfo {author} {\bibfnamefont {E.}~\bibnamefont {Rotenberg}}, \bibinfo {author} {\bibfnamefont {Q.}~\bibnamefont {Zhao}}, \bibinfo {author} {\bibfnamefont {J.~F.}\ \bibnamefont {Mitchell}}, \bibinfo {author} {\bibfnamefont {J.~W.}\ \bibnamefont {Allen}},\ and\ \bibinfo {author} {\bibfnamefont {B.~J.}\ \bibnamefont {Kim}},\ }\bibfield  {title} {\bibinfo {title} {Fermi arcs in a doped pseudospin$-1/2$ {H}eisenberg antiferromagnet},\ }\href {http://dx.doi.org/10.1126/science.1251151} {\bibfield  {journal} {\bibinfo  {journal} {Science}\ }\textbf {\bibinfo {volume} {345}},\ \bibinfo {pages} {187} (\bibinfo {year} {2014})}\BibitemShut {NoStop}%
\bibitem [{\citenamefont {Crawford}\ \emph {et~al.}(1994)\citenamefont {Crawford}, \citenamefont {Subramanian}, \citenamefont {Harlow}, \citenamefont {Fernandez-Baca}, \citenamefont {Wang},\ and\ \citenamefont {Johnston}}]{Crawford1994}%
  \BibitemOpen
  \bibfield  {author} {\bibinfo {author} {\bibfnamefont {M.~K.}\ \bibnamefont {Crawford}}, \bibinfo {author} {\bibfnamefont {M.~A.}\ \bibnamefont {Subramanian}}, \bibinfo {author} {\bibfnamefont {R.~L.}\ \bibnamefont {Harlow}}, \bibinfo {author} {\bibfnamefont {J.~A.}\ \bibnamefont {Fernandez-Baca}}, \bibinfo {author} {\bibfnamefont {Z.~R.}\ \bibnamefont {Wang}},\ and\ \bibinfo {author} {\bibfnamefont {D.~C.}\ \bibnamefont {Johnston}},\ }\bibfield  {title} {\bibinfo {title} {Structural and magnetic studies of {Sr$_{2}$IrO$_{4}$}},\ }\href {http://dx.doi.org/10.1103/PhysRevB.49.9198} {\bibfield  {journal} {\bibinfo  {journal} {Physical Review B}\ }\textbf {\bibinfo {volume} {49}},\ \bibinfo {pages} {9198} (\bibinfo {year} {1994})}\BibitemShut {NoStop}%
\bibitem [{\citenamefont {Ye}\ \emph {et~al.}(2013)\citenamefont {Ye}, \citenamefont {Chi}, \citenamefont {Chakoumakos}, \citenamefont {Fernandez-Baca}, \citenamefont {Qi},\ and\ \citenamefont {Cao}}]{Ye2013}%
  \BibitemOpen
  \bibfield  {author} {\bibinfo {author} {\bibfnamefont {F.}~\bibnamefont {Ye}}, \bibinfo {author} {\bibfnamefont {S.}~\bibnamefont {Chi}}, \bibinfo {author} {\bibfnamefont {B.~C.}\ \bibnamefont {Chakoumakos}}, \bibinfo {author} {\bibfnamefont {J.~A.}\ \bibnamefont {Fernandez-Baca}}, \bibinfo {author} {\bibfnamefont {T.}~\bibnamefont {Qi}},\ and\ \bibinfo {author} {\bibfnamefont {G.}~\bibnamefont {Cao}},\ }\bibfield  {title} {\bibinfo {title} {Magnetic and crystal structures of {Sr$_{2}$IrO$_{4}$}: A neutron diffraction study},\ }\href {http://dx.doi.org/10.1103/PhysRevB.87.140406} {\bibfield  {journal} {\bibinfo  {journal} {Physical Review B}\ }\textbf {\bibinfo {volume} {87}},\ \bibinfo {pages} {140406(R)} (\bibinfo {year} {2013})}\BibitemShut {NoStop}%
\bibitem [{\citenamefont {Shen}\ \emph {et~al.}(2004)\citenamefont {Shen}, \citenamefont {Ronning}, \citenamefont {Lu}, \citenamefont {Lee}, \citenamefont {Ingle}, \citenamefont {Meevasana}, \citenamefont {Baumberger}, \citenamefont {Damascelli}, \citenamefont {Armitage}, \citenamefont {Miller}, \citenamefont {Kohsaka}, \citenamefont {Azuma}, \citenamefont {Takano}, \citenamefont {Takagi},\ and\ \citenamefont {Shen}}]{Shen2004}%
  \BibitemOpen
  \bibfield  {author} {\bibinfo {author} {\bibfnamefont {K.~M.}\ \bibnamefont {Shen}}, \bibinfo {author} {\bibfnamefont {F.}~\bibnamefont {Ronning}}, \bibinfo {author} {\bibfnamefont {D.~H.}\ \bibnamefont {Lu}}, \bibinfo {author} {\bibfnamefont {W.~S.}\ \bibnamefont {Lee}}, \bibinfo {author} {\bibfnamefont {N.~J.~C.}\ \bibnamefont {Ingle}}, \bibinfo {author} {\bibfnamefont {W.}~\bibnamefont {Meevasana}}, \bibinfo {author} {\bibfnamefont {F.}~\bibnamefont {Baumberger}}, \bibinfo {author} {\bibfnamefont {A.}~\bibnamefont {Damascelli}}, \bibinfo {author} {\bibfnamefont {N.~P.}\ \bibnamefont {Armitage}}, \bibinfo {author} {\bibfnamefont {L.~L.}\ \bibnamefont {Miller}}, \bibinfo {author} {\bibfnamefont {Y.}~\bibnamefont {Kohsaka}}, \bibinfo {author} {\bibfnamefont {M.}~\bibnamefont {Azuma}}, \bibinfo {author} {\bibfnamefont {M.}~\bibnamefont {Takano}}, \bibinfo {author} {\bibfnamefont {H.}~\bibnamefont {Takagi}},\ and\ \bibinfo {author} {\bibfnamefont {Z.-X.}\ \bibnamefont {Shen}},\ }\bibfield  {title} {\bibinfo
  {title} {Missing quasiparticles and the chemical potential puzzle in the doping evolution of the cuprate superconductors},\ }\href {http://dx.doi.org/10.1103/PhysRevLett.93.267002} {\bibfield  {journal} {\bibinfo  {journal} {Physical Review Letters}\ }\textbf {\bibinfo {volume} {93}},\ \bibinfo {pages} {267002} (\bibinfo {year} {2004})}\BibitemShut {NoStop}%
\bibitem [{\citenamefont {Fournier}\ \emph {et~al.}(2010)\citenamefont {Fournier}, \citenamefont {Levy}, \citenamefont {Pennec}, \citenamefont {McChesney}, \citenamefont {Bostwick}, \citenamefont {Rotenberg}, \citenamefont {Liang}, \citenamefont {Hardy}, \citenamefont {Bonn}, \citenamefont {Elfimov},\ and\ \citenamefont {Damascelli}}]{Fournier2010}%
  \BibitemOpen
  \bibfield  {author} {\bibinfo {author} {\bibfnamefont {D.}~\bibnamefont {Fournier}}, \bibinfo {author} {\bibfnamefont {G.}~\bibnamefont {Levy}}, \bibinfo {author} {\bibfnamefont {Y.}~\bibnamefont {Pennec}}, \bibinfo {author} {\bibfnamefont {J.~L.}\ \bibnamefont {McChesney}}, \bibinfo {author} {\bibfnamefont {A.}~\bibnamefont {Bostwick}}, \bibinfo {author} {\bibfnamefont {E.}~\bibnamefont {Rotenberg}}, \bibinfo {author} {\bibfnamefont {R.}~\bibnamefont {Liang}}, \bibinfo {author} {\bibfnamefont {W.~N.}\ \bibnamefont {Hardy}}, \bibinfo {author} {\bibfnamefont {D.~A.}\ \bibnamefont {Bonn}}, \bibinfo {author} {\bibfnamefont {I.~S.}\ \bibnamefont {Elfimov}},\ and\ \bibinfo {author} {\bibfnamefont {A.}~\bibnamefont {Damascelli}},\ }\bibfield  {title} {\bibinfo {title} {Loss of nodal quasiparticle integrity in underdoped {YBa$_{2}$Cu$_{3}$O$_{6+x}$}},\ }\href {http://dx.doi.org/10.1038/nphys1763} {\bibfield  {journal} {\bibinfo  {journal} {Nature Physics}\ }\textbf {\bibinfo {volume} {6}},\ \bibinfo {pages} {905}
  (\bibinfo {year} {2010})}\BibitemShut {NoStop}%
\bibitem [{\citenamefont {Moutenet}\ \emph {et~al.}(2018)\citenamefont {Moutenet}, \citenamefont {Georges},\ and\ \citenamefont {Ferrero}}]{Moutenet2018}%
  \BibitemOpen
  \bibfield  {author} {\bibinfo {author} {\bibfnamefont {A.}~\bibnamefont {Moutenet}}, \bibinfo {author} {\bibfnamefont {A.}~\bibnamefont {Georges}},\ and\ \bibinfo {author} {\bibfnamefont {M.}~\bibnamefont {Ferrero}},\ }\bibfield  {title} {\bibinfo {title} {Pseudogap and electronic structure of electron-doped {Sr$_{2}$IrO$_{4}$}},\ }\href {http://dx.doi.org/10.1103/PhysRevB.97.155109} {\bibfield  {journal} {\bibinfo  {journal} {Physical Review B}\ }\textbf {\bibinfo {volume} {97}},\ \bibinfo {pages} {155109} (\bibinfo {year} {2018})}\BibitemShut {NoStop}%
\bibitem [{\citenamefont {Lee}\ \emph {et~al.}(2007)\citenamefont {Lee}, \citenamefont {Vishik}, \citenamefont {Tanaka}, \citenamefont {Lu}, \citenamefont {Sasagawa}, \citenamefont {Nagaosa}, \citenamefont {Devereaux}, \citenamefont {Hussain},\ and\ \citenamefont {Shen}}]{Lee2007}%
  \BibitemOpen
  \bibfield  {author} {\bibinfo {author} {\bibfnamefont {W.~S.}\ \bibnamefont {Lee}}, \bibinfo {author} {\bibfnamefont {I.~M.}\ \bibnamefont {Vishik}}, \bibinfo {author} {\bibfnamefont {K.}~\bibnamefont {Tanaka}}, \bibinfo {author} {\bibfnamefont {D.~H.}\ \bibnamefont {Lu}}, \bibinfo {author} {\bibfnamefont {T.}~\bibnamefont {Sasagawa}}, \bibinfo {author} {\bibfnamefont {N.}~\bibnamefont {Nagaosa}}, \bibinfo {author} {\bibfnamefont {T.~P.}\ \bibnamefont {Devereaux}}, \bibinfo {author} {\bibfnamefont {Z.}~\bibnamefont {Hussain}},\ and\ \bibinfo {author} {\bibfnamefont {Z.~X.}\ \bibnamefont {Shen}},\ }\bibfield  {title} {\bibinfo {title} {Abrupt onset of a second energy gap at the superconducting transition of underdoped bi2212},\ }\href {https://doi.org/10.1038/nature06219} {\bibfield  {journal} {\bibinfo  {journal} {Nature}\ }\textbf {\bibinfo {volume} {450}},\ \bibinfo {pages} {81} (\bibinfo {year} {2007})}\BibitemShut {NoStop}%
\bibitem [{\citenamefont {McKeown~Walker}\ \emph {et~al.}(2015)\citenamefont {McKeown~Walker}, \citenamefont {Bruno}, \citenamefont {Wang}, \citenamefont {de~la Torre}, \citenamefont {Ricc{\'o}}, \citenamefont {Tamai}, \citenamefont {Kim}, \citenamefont {Hoesch}, \citenamefont {Shi}, \citenamefont {Bahramy}, \citenamefont {King},\ and\ \citenamefont {Baumberger}}]{McKeownWalker2015}%
  \BibitemOpen
  \bibfield  {author} {\bibinfo {author} {\bibfnamefont {S.}~\bibnamefont {McKeown~Walker}}, \bibinfo {author} {\bibfnamefont {F.~Y.}\ \bibnamefont {Bruno}}, \bibinfo {author} {\bibfnamefont {Z.}~\bibnamefont {Wang}}, \bibinfo {author} {\bibfnamefont {A.}~\bibnamefont {de~la Torre}}, \bibinfo {author} {\bibfnamefont {S.}~\bibnamefont {Ricc{\'o}}}, \bibinfo {author} {\bibfnamefont {A.}~\bibnamefont {Tamai}}, \bibinfo {author} {\bibfnamefont {T.~K.}\ \bibnamefont {Kim}}, \bibinfo {author} {\bibfnamefont {M.}~\bibnamefont {Hoesch}}, \bibinfo {author} {\bibfnamefont {M.}~\bibnamefont {Shi}}, \bibinfo {author} {\bibfnamefont {M.~S.}\ \bibnamefont {Bahramy}}, \bibinfo {author} {\bibfnamefont {P.~D.~C.}\ \bibnamefont {King}},\ and\ \bibinfo {author} {\bibfnamefont {F.}~\bibnamefont {Baumberger}},\ }\bibfield  {title} {\bibinfo {title} {{Carrier-Density Control of the SrTiO$_3$ (001) Surface 2D Electron Gas studied by ARPES}},\ }\href {https://doi.org/https://doi.org/10.1002/adma.201501556} {\bibfield  {journal}
  {\bibinfo  {journal} {Advanced Materials}\ }\textbf {\bibinfo {volume} {27}},\ \bibinfo {pages} {3894} (\bibinfo {year} {2015})}\BibitemShut {NoStop}%
\bibitem [{\citenamefont {Chen}\ \emph {et~al.}(2019)\citenamefont {Chen}, \citenamefont {Hashimoto}, \citenamefont {He}, \citenamefont {Song}, \citenamefont {Xu}, \citenamefont {He}, \citenamefont {Devereaux}, \citenamefont {Eisaki}, \citenamefont {Lu}, \citenamefont {Zaanen},\ and\ \citenamefont {Shen}}]{Chen2019}%
  \BibitemOpen
  \bibfield  {author} {\bibinfo {author} {\bibfnamefont {S.-D.}\ \bibnamefont {Chen}}, \bibinfo {author} {\bibfnamefont {M.}~\bibnamefont {Hashimoto}}, \bibinfo {author} {\bibfnamefont {Y.}~\bibnamefont {He}}, \bibinfo {author} {\bibfnamefont {D.}~\bibnamefont {Song}}, \bibinfo {author} {\bibfnamefont {K.-J.}\ \bibnamefont {Xu}}, \bibinfo {author} {\bibfnamefont {J.-F.}\ \bibnamefont {He}}, \bibinfo {author} {\bibfnamefont {T.~P.}\ \bibnamefont {Devereaux}}, \bibinfo {author} {\bibfnamefont {H.}~\bibnamefont {Eisaki}}, \bibinfo {author} {\bibfnamefont {D.-H.}\ \bibnamefont {Lu}}, \bibinfo {author} {\bibfnamefont {J.}~\bibnamefont {Zaanen}},\ and\ \bibinfo {author} {\bibfnamefont {Z.-X.}\ \bibnamefont {Shen}},\ }\bibfield  {title} {\bibinfo {title} {Incoherent strange metal sharply bounded by a critical doping in bi2212},\ }\href {https://doi.org/10.1126/science.aaw8850} {\bibfield  {journal} {\bibinfo  {journal} {Science}\ }\textbf {\bibinfo {volume} {366}},\ \bibinfo {pages} {1099} (\bibinfo {year}
  {2019})}\BibitemShut {NoStop}%
\bibitem [{\citenamefont {Kim}\ \emph {et~al.}(2015)\citenamefont {Kim}, \citenamefont {Sung}, \citenamefont {Denlinger},\ and\ \citenamefont {Kim}}]{Kim2015}%
  \BibitemOpen
  \bibfield  {author} {\bibinfo {author} {\bibfnamefont {Y.~K.}\ \bibnamefont {Kim}}, \bibinfo {author} {\bibfnamefont {N.~H.}\ \bibnamefont {Sung}}, \bibinfo {author} {\bibfnamefont {J.~D.}\ \bibnamefont {Denlinger}},\ and\ \bibinfo {author} {\bibfnamefont {B.~J.}\ \bibnamefont {Kim}},\ }\bibfield  {title} {\bibinfo {title} {Observation of a d-wave gap in electron-doped {Sr$_{2}$IrO$_{4}$}},\ }\href {http://dx.doi.org/10.1038/nphys3503} {\bibfield  {journal} {\bibinfo  {journal} {Nature Physics}\ }\textbf {\bibinfo {volume} {12}},\ \bibinfo {pages} {37} (\bibinfo {year} {2015})}\BibitemShut {NoStop}%
\bibitem [{\citenamefont {Yan}\ \emph {et~al.}(2015)\citenamefont {Yan}, \citenamefont {Ren}, \citenamefont {Xu}, \citenamefont {Xie}, \citenamefont {Tao}, \citenamefont {Choi}, \citenamefont {Lee}, \citenamefont {Choi}, \citenamefont {Zhang},\ and\ \citenamefont {Feng}}]{Yan2015}%
  \BibitemOpen
  \bibfield  {author} {\bibinfo {author} {\bibfnamefont {Y.}~\bibnamefont {Yan}}, \bibinfo {author} {\bibfnamefont {M.}~\bibnamefont {Ren}}, \bibinfo {author} {\bibfnamefont {H.}~\bibnamefont {Xu}}, \bibinfo {author} {\bibfnamefont {B.}~\bibnamefont {Xie}}, \bibinfo {author} {\bibfnamefont {R.}~\bibnamefont {Tao}}, \bibinfo {author} {\bibfnamefont {H.}~\bibnamefont {Choi}}, \bibinfo {author} {\bibfnamefont {N.}~\bibnamefont {Lee}}, \bibinfo {author} {\bibfnamefont {Y.}~\bibnamefont {Choi}}, \bibinfo {author} {\bibfnamefont {T.}~\bibnamefont {Zhang}},\ and\ \bibinfo {author} {\bibfnamefont {D.}~\bibnamefont {Feng}},\ }\bibfield  {title} {\bibinfo {title} {Electron-doped {Sr$_{2}$IrO$_{4}$}: An analogue of hole-doped cuprate superconductors demonstrated by scanning tunneling microscopy},\ }\href {http://dx.doi.org/10.1103/PhysRevX.5.041018} {\bibfield  {journal} {\bibinfo  {journal} {Physical Review X}\ }\textbf {\bibinfo {volume} {5}},\ \bibinfo {pages} {041018} (\bibinfo {year} {2015})}\BibitemShut {NoStop}%
\bibitem [{\citenamefont {Yoshida}\ \emph {et~al.}(2009)\citenamefont {Yoshida}, \citenamefont {Hashimoto}, \citenamefont {Ideta}, \citenamefont {Fujimori}, \citenamefont {Tanaka}, \citenamefont {Mannella}, \citenamefont {Hussain}, \citenamefont {Shen}, \citenamefont {Kubota}, \citenamefont {Ono}, \citenamefont {Komiya}, \citenamefont {Ando}, \citenamefont {Eisaki},\ and\ \citenamefont {Uchida}}]{Yoshida2009}%
  \BibitemOpen
  \bibfield  {author} {\bibinfo {author} {\bibfnamefont {T.}~\bibnamefont {Yoshida}}, \bibinfo {author} {\bibfnamefont {M.}~\bibnamefont {Hashimoto}}, \bibinfo {author} {\bibfnamefont {S.}~\bibnamefont {Ideta}}, \bibinfo {author} {\bibfnamefont {A.}~\bibnamefont {Fujimori}}, \bibinfo {author} {\bibfnamefont {K.}~\bibnamefont {Tanaka}}, \bibinfo {author} {\bibfnamefont {N.}~\bibnamefont {Mannella}}, \bibinfo {author} {\bibfnamefont {Z.}~\bibnamefont {Hussain}}, \bibinfo {author} {\bibfnamefont {Z.-X.}\ \bibnamefont {Shen}}, \bibinfo {author} {\bibfnamefont {M.}~\bibnamefont {Kubota}}, \bibinfo {author} {\bibfnamefont {K.}~\bibnamefont {Ono}}, \bibinfo {author} {\bibfnamefont {S.}~\bibnamefont {Komiya}}, \bibinfo {author} {\bibfnamefont {Y.}~\bibnamefont {Ando}}, \bibinfo {author} {\bibfnamefont {H.}~\bibnamefont {Eisaki}},\ and\ \bibinfo {author} {\bibfnamefont {S.}~\bibnamefont {Uchida}},\ }\bibfield  {title} {\bibinfo {title} {Universal versus material-dependent two-gap behaviors of the high-tc cuprate
  superconductors: Angle-resolved photoemission study of {La$_{2-x}$Sr$_{x}$CuO$_{4}$}},\ }\href {http://dx.doi.org/10.1103/PhysRevLett.103.037004} {\bibfield  {journal} {\bibinfo  {journal} {Physical Review Letters}\ }\textbf {\bibinfo {volume} {103}},\ \bibinfo {pages} {037004} (\bibinfo {year} {2009})}\BibitemShut {NoStop}%
\bibitem [{\citenamefont {Cyr-Choini{\`e}re}\ \emph {et~al.}(2018)\citenamefont {Cyr-Choini{\`e}re}, \citenamefont {Daou}, \citenamefont {Lalibert{\'e}}, \citenamefont {Collignon}, \citenamefont {Badoux}, \citenamefont {LeBoeuf}, \citenamefont {Chang}, \citenamefont {Ramshaw}, \citenamefont {Bonn}, \citenamefont {Hardy}, \citenamefont {Liang}, \citenamefont {Yan}, \citenamefont {Cheng}, \citenamefont {Zhou}, \citenamefont {Goodenough}, \citenamefont {Pyon}, \citenamefont {Takayama}, \citenamefont {Takagi}, \citenamefont {Doiron-Leyraud},\ and\ \citenamefont {Taillefer}}]{CyrChoinire2018}%
  \BibitemOpen
  \bibfield  {author} {\bibinfo {author} {\bibfnamefont {O.}~\bibnamefont {Cyr-Choini{\`e}re}}, \bibinfo {author} {\bibfnamefont {R.}~\bibnamefont {Daou}}, \bibinfo {author} {\bibfnamefont {F.}~\bibnamefont {Lalibert{\'e}}}, \bibinfo {author} {\bibfnamefont {C.}~\bibnamefont {Collignon}}, \bibinfo {author} {\bibfnamefont {S.}~\bibnamefont {Badoux}}, \bibinfo {author} {\bibfnamefont {D.}~\bibnamefont {LeBoeuf}}, \bibinfo {author} {\bibfnamefont {J.}~\bibnamefont {Chang}}, \bibinfo {author} {\bibfnamefont {B.~J.}\ \bibnamefont {Ramshaw}}, \bibinfo {author} {\bibfnamefont {D.~A.}\ \bibnamefont {Bonn}}, \bibinfo {author} {\bibfnamefont {W.~N.}\ \bibnamefont {Hardy}}, \bibinfo {author} {\bibfnamefont {R.}~\bibnamefont {Liang}}, \bibinfo {author} {\bibfnamefont {J.-Q.}\ \bibnamefont {Yan}}, \bibinfo {author} {\bibfnamefont {J.-G.}\ \bibnamefont {Cheng}}, \bibinfo {author} {\bibfnamefont {J.-S.}\ \bibnamefont {Zhou}}, \bibinfo {author} {\bibfnamefont {J.~B.}\ \bibnamefont {Goodenough}}, \bibinfo {author}
  {\bibfnamefont {S.}~\bibnamefont {Pyon}}, \bibinfo {author} {\bibfnamefont {T.}~\bibnamefont {Takayama}}, \bibinfo {author} {\bibfnamefont {H.}~\bibnamefont {Takagi}}, \bibinfo {author} {\bibfnamefont {N.}~\bibnamefont {Doiron-Leyraud}},\ and\ \bibinfo {author} {\bibfnamefont {L.}~\bibnamefont {Taillefer}},\ }\bibfield  {title} {\bibinfo {title} {Pseudogap temperature ${T}^{*}$ of cuprate superconductors from the nernst effect},\ }\href {http://dx.doi.org/10.1103/PhysRevB.97.064502} {\bibfield  {journal} {\bibinfo  {journal} {Physical Review B}\ }\textbf {\bibinfo {volume} {97}},\ \bibinfo {pages} {064502} (\bibinfo {year} {2018})}\BibitemShut {NoStop}%
\bibitem [{\citenamefont {Moon}\ \emph {et~al.}(2009)\citenamefont {Moon}, \citenamefont {Jin}, \citenamefont {Choi}, \citenamefont {Lee}, \citenamefont {Seo}, \citenamefont {Yu}, \citenamefont {Cao}, \citenamefont {Noh},\ and\ \citenamefont {Lee}}]{Moon2009}%
  \BibitemOpen
  \bibfield  {author} {\bibinfo {author} {\bibfnamefont {S.~J.}\ \bibnamefont {Moon}}, \bibinfo {author} {\bibfnamefont {H.}~\bibnamefont {Jin}}, \bibinfo {author} {\bibfnamefont {W.~S.}\ \bibnamefont {Choi}}, \bibinfo {author} {\bibfnamefont {J.~S.}\ \bibnamefont {Lee}}, \bibinfo {author} {\bibfnamefont {S.~S.~A.}\ \bibnamefont {Seo}}, \bibinfo {author} {\bibfnamefont {J.}~\bibnamefont {Yu}}, \bibinfo {author} {\bibfnamefont {G.}~\bibnamefont {Cao}}, \bibinfo {author} {\bibfnamefont {T.~W.}\ \bibnamefont {Noh}},\ and\ \bibinfo {author} {\bibfnamefont {Y.~S.}\ \bibnamefont {Lee}},\ }\bibfield  {title} {\bibinfo {title} {Temperature dependence of the electronic structure of the ${J}_{\mathrm{eff}}=1/2$ mott insulator {Sr$_{2}$IrO$_{4}$} studied by optical spectroscopy},\ }\href {http://dx.doi.org/10.1103/PhysRevB.80.195110} {\bibfield  {journal} {\bibinfo  {journal} {Physical Review B}\ }\textbf {\bibinfo {volume} {80}},\ \bibinfo {pages} {195110} (\bibinfo {year} {2009})}\BibitemShut {NoStop}%
\bibitem [{\citenamefont {Wang}\ \emph {et~al.}(2013)\citenamefont {Wang}, \citenamefont {Cao}, \citenamefont {Waugh}, \citenamefont {Park}, \citenamefont {Qi}, \citenamefont {Korneta}, \citenamefont {Cao},\ and\ \citenamefont {Dessau}}]{Wang2013}%
  \BibitemOpen
  \bibfield  {author} {\bibinfo {author} {\bibfnamefont {Q.}~\bibnamefont {Wang}}, \bibinfo {author} {\bibfnamefont {Y.}~\bibnamefont {Cao}}, \bibinfo {author} {\bibfnamefont {J.~A.}\ \bibnamefont {Waugh}}, \bibinfo {author} {\bibfnamefont {S.~R.}\ \bibnamefont {Park}}, \bibinfo {author} {\bibfnamefont {T.~F.}\ \bibnamefont {Qi}}, \bibinfo {author} {\bibfnamefont {O.~B.}\ \bibnamefont {Korneta}}, \bibinfo {author} {\bibfnamefont {G.}~\bibnamefont {Cao}},\ and\ \bibinfo {author} {\bibfnamefont {D.~S.}\ \bibnamefont {Dessau}},\ }\bibfield  {title} {\bibinfo {title} {Dimensionality-controlled mott transition and correlation effects in single-layer and bilayer perovskite iridates},\ }\href {http://dx.doi.org/10.1103/PhysRevB.87.245109} {\bibfield  {journal} {\bibinfo  {journal} {Physical Review B}\ }\textbf {\bibinfo {volume} {87}},\ \bibinfo {pages} {245109} (\bibinfo {year} {2013})}\BibitemShut {NoStop}%
\bibitem [{\citenamefont {Coldea}\ \emph {et~al.}(2001)\citenamefont {Coldea}, \citenamefont {Hayden}, \citenamefont {Aeppli}, \citenamefont {Perring}, \citenamefont {Frost}, \citenamefont {Mason}, \citenamefont {Cheong},\ and\ \citenamefont {Fisk}}]{Coldea2001}%
  \BibitemOpen
  \bibfield  {author} {\bibinfo {author} {\bibfnamefont {R.}~\bibnamefont {Coldea}}, \bibinfo {author} {\bibfnamefont {S.~M.}\ \bibnamefont {Hayden}}, \bibinfo {author} {\bibfnamefont {G.}~\bibnamefont {Aeppli}}, \bibinfo {author} {\bibfnamefont {T.~G.}\ \bibnamefont {Perring}}, \bibinfo {author} {\bibfnamefont {C.~D.}\ \bibnamefont {Frost}}, \bibinfo {author} {\bibfnamefont {T.~E.}\ \bibnamefont {Mason}}, \bibinfo {author} {\bibfnamefont {S.-W.}\ \bibnamefont {Cheong}},\ and\ \bibinfo {author} {\bibfnamefont {Z.}~\bibnamefont {Fisk}},\ }\bibfield  {title} {\bibinfo {title} {Spin waves and electronic interactions in {La$_{2}$CuO${4}$}},\ }\href {https://doi.org/10.1103/PhysRevLett.86.5377} {\bibfield  {journal} {\bibinfo  {journal} {Phys. Rev. Lett.}\ }\textbf {\bibinfo {volume} {86}},\ \bibinfo {pages} {5377} (\bibinfo {year} {2001})}\BibitemShut {NoStop}%
\bibitem [{\citenamefont {Gretarsson}\ \emph {et~al.}(2016)\citenamefont {Gretarsson}, \citenamefont {Sung}, \citenamefont {Porras}, \citenamefont {Bertinshaw}, \citenamefont {Dietl}, \citenamefont {Bruin}, \citenamefont {Bangura}, \citenamefont {Kim}, \citenamefont {Dinnebier}, \citenamefont {Kim}, \citenamefont {Al-Zein}, \citenamefont {Moretti~Sala}, \citenamefont {Krisch}, \citenamefont {Le~Tacon}, \citenamefont {Keimer},\ and\ \citenamefont {Kim}}]{Gretarsson2016}%
  \BibitemOpen
  \bibfield  {author} {\bibinfo {author} {\bibfnamefont {H.}~\bibnamefont {Gretarsson}}, \bibinfo {author} {\bibfnamefont {N.}~\bibnamefont {Sung}}, \bibinfo {author} {\bibfnamefont {J.}~\bibnamefont {Porras}}, \bibinfo {author} {\bibfnamefont {J.}~\bibnamefont {Bertinshaw}}, \bibinfo {author} {\bibfnamefont {C.}~\bibnamefont {Dietl}}, \bibinfo {author} {\bibfnamefont {J.~A.}\ \bibnamefont {Bruin}}, \bibinfo {author} {\bibfnamefont {A.}~\bibnamefont {Bangura}}, \bibinfo {author} {\bibfnamefont {Y.}~\bibnamefont {Kim}}, \bibinfo {author} {\bibfnamefont {R.}~\bibnamefont {Dinnebier}}, \bibinfo {author} {\bibfnamefont {J.}~\bibnamefont {Kim}}, \bibinfo {author} {\bibfnamefont {A.}~\bibnamefont {Al-Zein}}, \bibinfo {author} {\bibfnamefont {M.}~\bibnamefont {Moretti~Sala}}, \bibinfo {author} {\bibfnamefont {M.}~\bibnamefont {Krisch}}, \bibinfo {author} {\bibfnamefont {M.}~\bibnamefont {Le~Tacon}}, \bibinfo {author} {\bibfnamefont {B.}~\bibnamefont {Keimer}},\ and\ \bibinfo {author} {\bibfnamefont {B.}~\bibnamefont
  {Kim}},\ }\bibfield  {title} {\bibinfo {title} {Persistent paramagnons deep in the metallic phase of {Sr$_{2-x}$La$_{x}$IrO$_{4}$}},\ }\href {http://dx.doi.org/10.1103/PhysRevLett.117.107001} {\bibfield  {journal} {\bibinfo  {journal} {Physical Review Letters}\ }\textbf {\bibinfo {volume} {117}},\ \bibinfo {pages} {107001} (\bibinfo {year} {2016})}\BibitemShut {NoStop}%
\bibitem [{\citenamefont {Keimer}\ \emph {et~al.}(1992)\citenamefont {Keimer}, \citenamefont {Belk}, \citenamefont {Birgeneau}, \citenamefont {Cassanho}, \citenamefont {Chen}, \citenamefont {Greven}, \citenamefont {Kastner}, \citenamefont {Aharony}, \citenamefont {Endoh}, \citenamefont {Erwin},\ and\ \citenamefont {Shirane}}]{Keimer1992}%
  \BibitemOpen
  \bibfield  {author} {\bibinfo {author} {\bibfnamefont {B.}~\bibnamefont {Keimer}}, \bibinfo {author} {\bibfnamefont {N.}~\bibnamefont {Belk}}, \bibinfo {author} {\bibfnamefont {R.~J.}\ \bibnamefont {Birgeneau}}, \bibinfo {author} {\bibfnamefont {A.}~\bibnamefont {Cassanho}}, \bibinfo {author} {\bibfnamefont {C.~Y.}\ \bibnamefont {Chen}}, \bibinfo {author} {\bibfnamefont {M.}~\bibnamefont {Greven}}, \bibinfo {author} {\bibfnamefont {M.~A.}\ \bibnamefont {Kastner}}, \bibinfo {author} {\bibfnamefont {A.}~\bibnamefont {Aharony}}, \bibinfo {author} {\bibfnamefont {Y.}~\bibnamefont {Endoh}}, \bibinfo {author} {\bibfnamefont {R.~W.}\ \bibnamefont {Erwin}},\ and\ \bibinfo {author} {\bibfnamefont {G.}~\bibnamefont {Shirane}},\ }\bibfield  {title} {\bibinfo {title} {Magnetic excitations in pure, lightly doped, and weakly metallic {La$_{2}$CuO$_{4}$}},\ }\href {http://dx.doi.org/10.1103/PhysRevB.46.14034} {\bibfield  {journal} {\bibinfo  {journal} {Physical Review B}\ }\textbf {\bibinfo {volume} {46}},\ \bibinfo
  {pages} {14034} (\bibinfo {year} {1992})}\BibitemShut {NoStop}%
\bibitem [{\citenamefont {Drachuck}\ \emph {et~al.}(2014)\citenamefont {Drachuck}, \citenamefont {Razzoli}, \citenamefont {Bazalitski}, \citenamefont {Kanigel}, \citenamefont {Niedermayer}, \citenamefont {Shi},\ and\ \citenamefont {Keren}}]{Drachuck2014}%
  \BibitemOpen
  \bibfield  {author} {\bibinfo {author} {\bibfnamefont {G.}~\bibnamefont {Drachuck}}, \bibinfo {author} {\bibfnamefont {E.}~\bibnamefont {Razzoli}}, \bibinfo {author} {\bibfnamefont {G.}~\bibnamefont {Bazalitski}}, \bibinfo {author} {\bibfnamefont {A.}~\bibnamefont {Kanigel}}, \bibinfo {author} {\bibfnamefont {C.}~\bibnamefont {Niedermayer}}, \bibinfo {author} {\bibfnamefont {M.}~\bibnamefont {Shi}},\ and\ \bibinfo {author} {\bibfnamefont {A.}~\bibnamefont {Keren}},\ }\bibfield  {title} {\bibinfo {title} {Comprehensive study of the spin-charge interplay in antiferromagnetic {La$_{2-x}$Sr$_{x}$CuO$_{4}$}},\ }\href {http://dx.doi.org/10.1038/ncomms4390} {\bibfield  {journal} {\bibinfo  {journal} {Nature Communications}\ }\textbf {\bibinfo {volume} {5}},\ \bibinfo {pages} {3390} (\bibinfo {year} {2014})}\BibitemShut {NoStop}%
\bibitem [{\citenamefont {Kyung}\ \emph {et~al.}(2006)\citenamefont {Kyung}, \citenamefont {Kancharla}, \citenamefont {S{\'e}n{\'e}chal}, \citenamefont {Tremblay}, \citenamefont {Civelli},\ and\ \citenamefont {Kotliar}}]{Kyung2006}%
  \BibitemOpen
  \bibfield  {author} {\bibinfo {author} {\bibfnamefont {B.}~\bibnamefont {Kyung}}, \bibinfo {author} {\bibfnamefont {S.~S.}\ \bibnamefont {Kancharla}}, \bibinfo {author} {\bibfnamefont {D.}~\bibnamefont {S{\'e}n{\'e}chal}}, \bibinfo {author} {\bibfnamefont {A.-M.~S.}\ \bibnamefont {Tremblay}}, \bibinfo {author} {\bibfnamefont {M.}~\bibnamefont {Civelli}},\ and\ \bibinfo {author} {\bibfnamefont {G.}~\bibnamefont {Kotliar}},\ }\bibfield  {title} {\bibinfo {title} {Pseudogap induced by short-range spin correlations in a doped mott insulator},\ }\href {http://dx.doi.org/10.1103/PhysRevB.73.165114} {\bibfield  {journal} {\bibinfo  {journal} {Physical Review B}\ }\textbf {\bibinfo {volume} {73}},\ \bibinfo {pages} {165114} (\bibinfo {year} {2006})}\BibitemShut {NoStop}%
\bibitem [{\citenamefont {Gunnarsson}\ \emph {et~al.}(2015)\citenamefont {Gunnarsson}, \citenamefont {Sch\"{a}fer}, \citenamefont {LeBlanc}, \citenamefont {Gull}, \citenamefont {Merino}, \citenamefont {Sangiovanni}, \citenamefont {Rohringer},\ and\ \citenamefont {Toschi}}]{Gunnarsson2015}%
  \BibitemOpen
  \bibfield  {author} {\bibinfo {author} {\bibfnamefont {O.}~\bibnamefont {Gunnarsson}}, \bibinfo {author} {\bibfnamefont {T.}~\bibnamefont {Sch\"{a}fer}}, \bibinfo {author} {\bibfnamefont {J.}~\bibnamefont {LeBlanc}}, \bibinfo {author} {\bibfnamefont {E.}~\bibnamefont {Gull}}, \bibinfo {author} {\bibfnamefont {J.}~\bibnamefont {Merino}}, \bibinfo {author} {\bibfnamefont {G.}~\bibnamefont {Sangiovanni}}, \bibinfo {author} {\bibfnamefont {G.}~\bibnamefont {Rohringer}},\ and\ \bibinfo {author} {\bibfnamefont {A.}~\bibnamefont {Toschi}},\ }\bibfield  {title} {\bibinfo {title} {Fluctuation diagnostics of the electron self-energy: Origin of the pseudogap physics},\ }\href {http://dx.doi.org/10.1103/PhysRevLett.114.236402} {\bibfield  {journal} {\bibinfo  {journal} {Physical Review Letters}\ }\textbf {\bibinfo {volume} {114}},\ \bibinfo {pages} {236402} (\bibinfo {year} {2015})}\BibitemShut {NoStop}%
\bibitem [{\citenamefont {Wu}\ \emph {et~al.}(2017)\citenamefont {Wu}, \citenamefont {Ferrero}, \citenamefont {Georges},\ and\ \citenamefont {Kozik}}]{Wu2017}%
  \BibitemOpen
  \bibfield  {author} {\bibinfo {author} {\bibfnamefont {W.}~\bibnamefont {Wu}}, \bibinfo {author} {\bibfnamefont {M.}~\bibnamefont {Ferrero}}, \bibinfo {author} {\bibfnamefont {A.}~\bibnamefont {Georges}},\ and\ \bibinfo {author} {\bibfnamefont {E.}~\bibnamefont {Kozik}},\ }\bibfield  {title} {\bibinfo {title} {Controlling feynman diagrammatic expansions: Physical nature of the pseudogap in the two-dimensional hubbard model},\ }\href {http://dx.doi.org/10.1103/PhysRevB.96.041105} {\bibfield  {journal} {\bibinfo  {journal} {Physical Review B}\ }\textbf {\bibinfo {volume} {96}},\ \bibinfo {pages} {041105(R)} (\bibinfo {year} {2017})}\BibitemShut {NoStop}%
\bibitem [{\citenamefont {Wang}\ \emph {et~al.}(2015)\citenamefont {Wang}, \citenamefont {Yu},\ and\ \citenamefont {Li}}]{Wang2015}%
  \BibitemOpen
  \bibfield  {author} {\bibinfo {author} {\bibfnamefont {H.}~\bibnamefont {Wang}}, \bibinfo {author} {\bibfnamefont {S.-L.}\ \bibnamefont {Yu}},\ and\ \bibinfo {author} {\bibfnamefont {J.-X.}\ \bibnamefont {Li}},\ }\bibfield  {title} {\bibinfo {title} {Fermi arcs, pseudogap, and collective excitations in doped {Sr$_{2}$IrO$_{4}$}: A generalized fluctuation exchange study},\ }\href {http://dx.doi.org/10.1103/PhysRevB.91.165138} {\bibfield  {journal} {\bibinfo  {journal} {Physical Review B}\ }\textbf {\bibinfo {volume} {91}},\ \bibinfo {pages} {165138} (\bibinfo {year} {2015})}\BibitemShut {NoStop}%
\bibitem [{\citenamefont {{\v S}imkovic}\ \emph {et~al.}(2024)\citenamefont {{\v S}imkovic}, \citenamefont {Rossi}, \citenamefont {Georges},\ and\ \citenamefont {Ferrero}}]{Simkovic2024}%
  \BibitemOpen
  \bibfield  {author} {\bibinfo {author} {\bibfnamefont {F.}~\bibnamefont {{\v S}imkovic}}, \bibinfo {author} {\bibfnamefont {R.}~\bibnamefont {Rossi}}, \bibinfo {author} {\bibfnamefont {A.}~\bibnamefont {Georges}},\ and\ \bibinfo {author} {\bibfnamefont {M.}~\bibnamefont {Ferrero}},\ }\bibfield  {title} {\bibinfo {title} {{Origin and fate of the pseudogap in the doped Hubbard model}},\ }\href {https://doi.org/10.1126/science.ade9194} {\bibfield  {journal} {\bibinfo  {journal} {Science}\ }\textbf {\bibinfo {volume} {385}},\ \bibinfo {pages} {eade9194} (\bibinfo {year} {2024})}\BibitemShut {NoStop}%
\bibitem [{\citenamefont {Nelson}\ \emph {et~al.}(2020)\citenamefont {Nelson}, \citenamefont {Parzyck}, \citenamefont {Faeth}, \citenamefont {Kawasaki}, \citenamefont {Schlom},\ and\ \citenamefont {Shen}}]{Nelson2020}%
  \BibitemOpen
  \bibfield  {author} {\bibinfo {author} {\bibfnamefont {J.~N.}\ \bibnamefont {Nelson}}, \bibinfo {author} {\bibfnamefont {C.~T.}\ \bibnamefont {Parzyck}}, \bibinfo {author} {\bibfnamefont {B.~D.}\ \bibnamefont {Faeth}}, \bibinfo {author} {\bibfnamefont {J.~K.}\ \bibnamefont {Kawasaki}}, \bibinfo {author} {\bibfnamefont {D.~G.}\ \bibnamefont {Schlom}},\ and\ \bibinfo {author} {\bibfnamefont {K.~M.}\ \bibnamefont {Shen}},\ }\bibfield  {title} {\bibinfo {title} {Mott gap collapse in lightly hole-doped {Sr$_{2-x}$K$_{x}$IrO$_{4}$}},\ }\href {http://dx.doi.org/10.1038/s41467-020-16425-z} {\bibfield  {journal} {\bibinfo  {journal} {Nature Communications}\ }\textbf {\bibinfo {volume} {11}},\ \bibinfo {pages} {2597} (\bibinfo {year} {2020})}\BibitemShut {NoStop}%
\bibitem [{\citenamefont {Armitage}\ \emph {et~al.}(2010)\citenamefont {Armitage}, \citenamefont {Fournier},\ and\ \citenamefont {Greene}}]{Armitage2010}%
  \BibitemOpen
  \bibfield  {author} {\bibinfo {author} {\bibfnamefont {N.~P.}\ \bibnamefont {Armitage}}, \bibinfo {author} {\bibfnamefont {P.}~\bibnamefont {Fournier}},\ and\ \bibinfo {author} {\bibfnamefont {R.~L.}\ \bibnamefont {Greene}},\ }\bibfield  {title} {\bibinfo {title} {Progress and perspectives on electron-doped cuprates},\ }\href {https://doi.org/10.1103/RevModPhys.82.2421} {\bibfield  {journal} {\bibinfo  {journal} {Rev. Mod. Phys.}\ }\textbf {\bibinfo {volume} {82}},\ \bibinfo {pages} {2421} (\bibinfo {year} {2010})}\BibitemShut {NoStop}%
\bibitem [{\citenamefont {P\"{a}rschke}\ \emph {et~al.}(2017)\citenamefont {P\"{a}rschke}, \citenamefont {Wohlfeld}, \citenamefont {Foyevtsova},\ and\ \citenamefont {van~den Brink}}]{Parschke2017}%
  \BibitemOpen
  \bibfield  {author} {\bibinfo {author} {\bibfnamefont {E.~M.}\ \bibnamefont {P\"{a}rschke}}, \bibinfo {author} {\bibfnamefont {K.}~\bibnamefont {Wohlfeld}}, \bibinfo {author} {\bibfnamefont {K.}~\bibnamefont {Foyevtsova}},\ and\ \bibinfo {author} {\bibfnamefont {J.}~\bibnamefont {van~den Brink}},\ }\bibfield  {title} {\bibinfo {title} {Correlation induced electron-hole asymmetry in quasi- two-dimensional iridates},\ }\href {http://dx.doi.org/10.1038/s41467-017-00818-8} {\bibfield  {journal} {\bibinfo  {journal} {Nature Communications}\ }\textbf {\bibinfo {volume} {8}},\ \bibinfo {pages} {686} (\bibinfo {year} {2017})}\BibitemShut {NoStop}%
\bibitem [{dat()}]{data}%
  \BibitemOpen
  \href@noop {} {}\bibinfo {note} {The datasets analyzed during this study are available at the Yareta repository of the University of Geneva: \href{https://doi.org/10.26037/yareta:xkpk4bojqzfctdt2eb34kl64ie}{10.26037/yareta:xkpk4bojqzfctdt2eb34kl64ie}}\BibitemShut {NoStop}%
\end{thebibliography}
\end{document}